\documentclass[10pt]{iopart}

\usepackage{iopams}

\usepackage{graphicx}

\usepackage[position=top]{subcaption}
\usepackage{multirow}
\usepackage{longtable}

\usepackage{afterpage}
\usepackage{xtab}

\newcommand{\lvec}{\boldsymbol{l}}
\newcommand{\svec}{\boldsymbol{s}}
\newcommand{\etavec}{\boldsymbol{\eta}}
\newcommand{\exvec}{\mathbf{e}_{x}}
\newcommand{\eyvec}{\mathbf{e}_{y}}
\newcommand{\ezvec}{\mathbf{e}_{z}}

\begin{document}

\title[Rapid prediction of organisation in engineered corneal, glial and fibroblast tissues]{Rapid prediction of organisation in engineered corneal, glial and fibroblast tissues using machine learning and biophysical models}

\author{Allison E. Andrews$^{1}$, Hugh Dickinson$^{1}$, Caitriona O'Rourke$^{2}$, James B. Philips$^{2,3}$ and James P. Hague$^{1}$}

\address{$^1$School of Physical Science, The Open University, Walton Hall, Milton Keynes, MK7 6AA, United Kingdom}
\address{$^2$UCL Centre for Nerve Engineering, London WC1E 6BT, United Kingdom}
\address{$^3$Department of Pharmacology, UCL School of Pharmacy, London WC1N 1AX, United Kingdom}

\date{24th September 2026} 

\begin{abstract}
We present a machine learning approach for predicting the organisation of corneal, glial and fibroblast cells in 3D cultures used for tissue engineering. Our machine-learning-based method uses a generative adversarial network architecture called \texttt{pix2pix}, which we train using results from biophysical contractile network dipole orientation (CONDOR) simulations. In the following, we refer to the machine learning method as the RAPTOR (RApid Prediction of Tissue ORganisation) approach. A training data set containing a range of CONDOR simulations is created, covering a range of underlying model parameters. Predictions of the trained neural network are compared with cultured glial, corneal, and fibroblast tissues, with good agreements for both CONDOR and RAPTOR approaches. An approach is developed to determine CONDOR model parameters for specific tissues using both RAPTOR and CONDOR fits to tissue properties. RAPTOR outputs a variety of tissue properties, including cell densities, cell alignments and tension. RAPTOR yields predictions of tissue properties within fractions of a second. This speed makes it valuable for the design of tethered moulds for tissue growth.    
\end{abstract}

\ioptwocol

\section{Introduction}

Cultured tissues are complex and expensive to design, and tools to help this design process would be valuable. While cells self-organise in a bottom-up process, design of moulds and scaffolds for tissue growth is easiest from a top-down perspective. The costs associated with growing tissues for experimental mould design can be  significant, and individual tissues may fail during the growth process further increasing expenditure. Tools to carry out simulations and make predictions are important to assist with this design process.

The self-organisation of cells is key to tissue function. The cells and matrix of corneal cells organise to create its optical and mechanical properties \cite{gouveia2017a}. Glial cells organise to support nerve function \cite{orourke2015}. Highly aligned fibroblasts are key to the healing process \cite{werner2007a}. Highly-aligned myocytes form into muscles \cite{grosberg2011a,shimizu2009}. Recently, cultured tissues have been developed that emulate the function and structure of naturally occurring tissues in humans and other animals, with applications including regenerative medicine \cite{bajaj2014a}, cultivated meats \cite{benarye2019a} and assays for pharmacological testing  \cite{weinhart2019a,jensen2018a}. Improved control over cell organisation in cultivated tissues could lead to benefits for e.g. nerve tissue, engineered tendon, skeletal muscle and dentine \cite{lu2021a}. For pharmaceutical testing, cultured tissues with cell organisations that closely replicate naturally occurring tissues are needed, as this can affect gene expression \cite{jensen2020a}.  For cultivated meat, cell function is less important, however the overall structure of myocytes, adipocytes and matrix in the tissue has important implications for taste and texture \cite{benarye2019a}.

The contractile network dipole orientation (CONDOR) method is a useful tool for the design of scaffolds for tissue growth \cite{hague2019a,andrews2023,hague2023a}. For scoping out designs, it offers a cheaper and faster alternative to lab work as it is self-analysing (i.e. the cell density, tissue shape, cell orientation and internal forces can be read directly from the simulation results) rather than needing extensive microscopy, and simulations can be run as a massively parallel process with scaffolds and moulds designed automatically \cite{andrews2023,hague2023a}.  The size and shape of cultivated tissues depends on the application. For nerve, long thin cylinders are needed, whereas for skin or cornea a square or round sheet might be preferred. The target structure of the tissue may also vary. For example, cultured tissues for neural engineering may need to maximise total alignment (i.e. minimise the volume of any unaligned area) and have even and consistent alignment throughout the material. Other aspects of the tissue may also need optimising. For example, when tissues are grown in moulds, residual tension might be problematic if the tissue needs to be removed from the mould. Furthermore, there may be other practical issues providing constraints on mould designs, such as the material used to make the mould and the mould manufacture technique, both of which can limit its possible dimensions. Also moulds need to be sterile (or easily sterilised) and suitable for immersion in media in an incubator. All these tissue properties and constraints can be explored within CONDOR simulations. See Fig. \ref{fig:schematic} for an example of a contractile network. Details of the CONDOR model are introduced in Section \ref{sec:condor}.

Machine learning (ML) techniques based on CONDOR can speed up the prediction of self-organisation in tissues, opening the possibility of real-time design. As proof of concept, an implementation of the \texttt{pix2pix} generative adversarial network architecture \cite{isola2017a} was trained using a single cell-matrix interaction parameter consistent with 3D cultures of glial cells \cite{andrews2023}. This technique was denoted CONDOR-ML. Given an input image representing the shape of a tethered mould, the resulting CONDOR-ML approach makes very rapid predictions of cell organisation in cultured glial tissues \cite{andrews2023}. A limitation of CONDOR-ML is that it has fixed interaction and ECM parameters and these can vary for other tissue and 3D culture types. 

Tethering refers to the integration of part of the hydrogel with a rigid structure. For the cases modelled in this paper, the hydrogels were cast into moulds with posts at each end. These posts provide fixed anchor points that the hydrogel surrounds, tethering it at those points. Other methods for tethering have been described previously, for example using a stiff mesh embedded within the mould rather than using posts (see Ref. \cite{phillips2011}). Tethering of hydrogels is a tissue engineering technique used to harness and direct endogenous cellular contraction. Cells within gels form integrin-mediated attachments with the collagen fibrils in the hydrogel and the resulting endogenous contractile forces exerted by the cells on the surrounding hydrogel cause an overall shrinkage of the cellular material. By tethering the hydrogel, e.g. at opposite ends of a rectangular mould, the strain generated by the cells causes tension to build up parallel to the long axis of the hydrogel. Without the tethers, the tissue would contract, but only weak forces and limited self-organisation would be found.

The purpose of this paper is to introduce and validate machine learning tools for the rapid prediction of tissue orientation (RAPTOR) for a variety of 3D tissue cultures. RAPTOR extends CONDOR-ML by including the possibility to change the underlying CONDOR parameters (i.e. to simulate tissues with a range of cell and matrix types). As part of a mould design process, the method could be useful to find ways to grow highly aligned tissues, to minimise tissue tension and also determine suitable matrix properties. Comparison is made with archive data for cultured tissue made from glial cells, corneal stromal cells, dermal fibroblasts, tenocytes and myoblasts. In this paper, we will focus on tissues grown in tethered cell-laden hydrogels (hydrogels such as collagen containing cells), one of several strategies for culturing tissues  \cite{bajaj2014a}. 

The relation between CONDOR, CONDOR-ML and RAPTOR is as follows. CONDOR is a microscopic biophysical approach for predicting the self-organisation of cells in tissues, and the resulting deformations to the tissue. Parameters represent the strength of interactions between cells and ECM, and the springs of a contractile network represent the ECM (Fig. \ref{fig:schematic}) \cite{hague2019a}. CONDOR-ML and RAPTOR are both \texttt{pix2pix} based approaches for predicting tissue organisation that are trained on output from CONDOR. CONDOR-ML is trained using a single specific set of cell-ECM and ECM parameters \cite{andrews2023}. RAPTOR is trained using a range of cell-ECM and matrix parameters, and can therefore be used to make predictions for a wider range of cell and matrix types.

This paper is organised as follows. In Sec. \ref{sec:methods} the methodology of the CONDOR model, details of the training data set and the RAPTOR approach are provided. In Sec. \ref{sec:results} validation of the RAPTOR approach is carried out. Finally, a summary and outlook is given in Sec. \ref{sec:summary}.

\section{Methods}
\label{sec:methods}

\subsection{Contractile Network Dipole Orientation Model}
\label{sec:condor}

The contractile network dipole orientation (CONDOR) method determines the self-organisation of polarised cells within tissues. Polarised cells are elongated with approximate order 2 rotational symmetry. Here CONDOR is used to find the positional arrangement and alignment of such cells, and the overall tissue shape. CONDOR can also be used to estimate the tension inside a tissue. CONDOR is particularly suitable for cell laden hydrogels, where cell density is sufficiently low that the cells are not closely packed, as is the case here. The methodology of CONDOR has been described in detail in Ref. \cite{hague2019a} so we provide a summary here.

The model comprises cells interacting through a contractile network of bonds. The minimum energy state of the  CONDOR model, determined using simulated annealing, can be used to find the self-organisation of cells in tissues \cite{hague2019a}.

\begin{figure}
\includegraphics[width=0.4\textwidth]{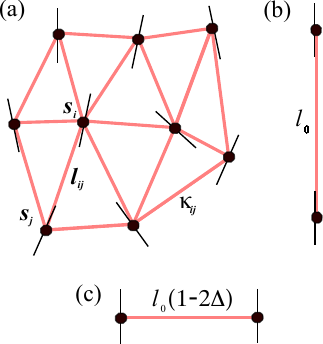}
    \caption{(a) Schematic of the CONDOR model. Cell orientations are represented using $\svec$ and the displacement between them by $\lvec$. The contractile network of springs (bonds) is shown as red lines. Bonds have a spring constant $\kappa_{ij}$. Only nearest neighbour bonds are shown in the figure, but we include bonds up to 4th nearest neighbour within the model. (b) When cells are aligned with bonds, the length is unmodified. (c) When cells are perpendicular to bonds, the length is reduced by $2\Delta$. In general, the length is reduced by $\Delta(\sin^2\theta_{1}+\sin^2\theta_{2})$ where $\theta_{1}$ and $\theta_{2}$ are the angles between the bond and the cell orientations at each end of the bond.}
    \label{fig:schematic}
\end{figure}

Cells generate active force dipoles within the tissue. A force dipole is characterised by opposite forces pulling inwards perpendicular to the dipole. These can be represented through $l'_{ij}$, the modification to the equilibrium bond length $l_{0}$, leading to cell-matrix interaction,
\begin{equation}
l'_{ij} = l_{0}\left(1-\frac{\Delta}{2}\left(2-|\hat{\lvec}_{ij}\cdot\svec_{i}|^2 -|\hat{\lvec}_{ij}\cdot\svec_{j}|^2\right) \right),
\label{eqn:interaction}
\end{equation}
By rewriting the scalar products, it can be seen that the length is reduced by $\Delta(\sin^2\theta_{1}+\sin^2\theta_{2})/2$ where $\theta_{1}$ and $\theta_{2}$ are the angles between the bond and the cell orientations at each end of the bond. Thus there is no reduction for cells directed along bonds and a maximum reduction for cells angled perpendicular to bonds.  A schematic figure showing the components of the model can be found in Fig. \ref{fig:schematic}, showing the elements of the model: the contractile network (as red lines), and force dipoles (as black lines at vertices of the network).

The total energy, $E$, of the system of contractile network and force dipoles (sum on bonds) within the CONDOR model is given by,
\begin{equation} 
E = \sum_{i<j} \frac{\kappa_{0}\bar{\kappa}_{ij}}{2}\left( |\lvec_{ij}| - l'_{ij}\right)^2.
\label{eqn:hamiltonian}
\end{equation}

The model is scale independent, with all results given in terms of dimensionless quantities. 
$\lvec_{ij}$ is the displacement between the centres of cells $i$ and $j$. The strength of the cell-matrix interaction is provided by the dimensionless number, $\Delta$. The dimensionless spring constant, $\bar{\kappa}_{ij}$, can be varied to modify the properties of the matrix. The orientations of cells and bonds are represented by unitary vectors $\svec$ and $\hat{\lvec}$ respectively. Since cells are represented as force dipoles, $\svec$ represents the orientation of the force dipole as well as the cell. The nearest-neighbour spring constant, $\kappa_{0}$, sets the energy scale. $\overline{\kappa}_{\rm 4N}$ and $\overline{\kappa}_{\rm 4N}$, are the next-next-nearest and next-next-next-nearest neighbour spring constants, respectively. Initially, cells are arranged on a face centre cubic (FCC) lattice. We use an FCC lattice to ensure that there is a suitable shear modulus even if there are just nearest neighbour interactions. Details of the bulk and shear moduli for the spring network can be found in \ref{sec:appendixa}. Tension in the bond is given by $T_{ij} = \kappa_0\kappa_{ij}l'_{ij}$. Simulation predictions are invariant with regard to choice of $l_{0}$. The model is expected to be most reliable for cases where the cell density is not too high (as is the case with most artificial cellular hydrogels). For systems with tight junctions, such as epithelium, a different model would be needed.

Equation \ref{eqn:interaction} has the order 2 rotational symmetry associated with the cells: a 180$^{o}$ rotation of $\svec_{i}$ results in the same system energy. The modification of the length is consistent with force dipoles, where cells pull on matrix perpendicular to their orientation. We note that while the model is scale invariant, the value of $\Delta$ should increase with cell density, since in that case there are more cells per unit volume to pull on the matrix.

The length scales associated with $l_0$ broadly correspond to the distance between cells or groups of cells. A typical mould has around 7000 cells, which is about 80 cells across. A typical mould side length is around 3mm. Thus the distance between cells, $l_0$ in the model, is approximately 30 microns. So, depending on the cell sizes, $l_0$ could correspond to a single cell or a group of cells. In the latter case, the force dipole would represent the effects of cell-matrix interactions due to a cluster of several cells.

The lowest energy configuration of Eq. \ref{eqn:hamiltonian} was determined with simulated annealing. For the CONDOR results the orientation of cells and their positions were changed on each of $\eta = N_{\rm cells} \times 10^{6}$ total iterations (total cell number is $N_{\rm cells}$). The number of updates is selected to be proportional to $N_{\rm cells}$ so that cells are visited the same number of times for systems of different sizes. The factor $10^{6}$ is selected as the results have been shown to be reliably converged for this anneal schedule by using multiple random seeds to initialise the random number generator used in the Metropolis algorithm underpinning simulated annealing (see error bars on Fig. 4 of Ref. \cite{hague2019a} - note that the improved run time here is due to code optimisations). Changes were accepted according to the Metropolis condition, where the probability of acceptance is,
\begin{equation}
    P({\rm accept}) = {\rm min}\left\{\exp\left(-\Delta E / {\mathcal{T}}\right),1\right\}
\end{equation}
with the anneal parameter, ${\mathcal{T}}$ reduced in an exponential schedule by a factor ${\mathcal{T}}_{\rm init}/{\mathcal{T}}_{\rm final}=10^{7}$ from an initial anneal parameter of ${\mathcal{T}}_{\rm init}/\kappa_{0}l_{0}^2=0.0625$. ${\mathcal{T}}_{\rm final}$ is the final anneal parameter. By performing Metropolis updates in this way, the parameter, ${\mathcal{T}}$, dictates the size of energy fluctuations - by reducing ${\mathcal{T}}$ sufficiently slowly, the minimum energy configuration is guaranteed to be reached. 

The exponential schedule is defined by, ${\mathcal{T}}_{\theta+1}=\epsilon {\mathcal{T}}_{\theta}$ 
where $\theta$ is the iteration number and $\epsilon=\exp\left(\ln{\mathcal{T}}_{\rm init}
-\ln{\mathcal{T}}_{\rm final}\right)/\eta$. 
For the mould sizes selected, this schedule leads to well optimised cell and matrix organisations. Each cell that moves outside the mould is subject to a large penalty. 

It takes approximately 24-36 core hours to run the CONDOR simulations for a mould with $\sim 10000$ cells. Other uses of CONDOR can be found in Ref. \cite{hague2019a,andrews2023,hague2023a}. 

\subsection{CONDOR Data}
\label{sec:trainingdata}

We use the CONDOR simulation to construct training and validation data sets. These were generated by automatically constructing random mould designs and tether layouts that are then used as inputs to CONDOR simulations, using the methods described in Section 2.1 of Ref.~\cite{andrews2023} (where full details of the process for constructing random mould designs can be found). In this work, the training data set additionally includes varying simulation parameters, namely $\Delta$, $\overline{\kappa}_{\rm 3N}$ and $\overline{\kappa}_{\rm 4N}$ (note $\overline{\kappa}_{\rm NN}=1$ throughout); these parameters affect the overall contractile behavior of simulated tissue. $\Delta$ controls the overall strength of contraction between simulated cells, with higher values leading to greater overall contraction and compression of the cell matrix. $\overline{\kappa}_{\rm 3N}$ and $\overline{\kappa}_{\rm 4N}$ are spring constants that limit contraction between non-adjacent neighboring cells (next-next-nearest for $\overline{\kappa}_{\rm 3N}$, and next-next-next-nearest for $\overline{\kappa}_{\rm 4N}$), i.e. they increase the bulk and shear modulus. Reducing either spring constant allows for the possibility of increased overall contraction, though contraction is most sensitive to variations in $\Delta$; spring constants also have subtle effects on the final contracted cell configurations, particularly with contracted shape (for example, lower spring constants lead to greater curvature in the final cell matrix boundary). By allowing these parameters to vary within the training data, they can be used to modulate predictions produced by the final trained model, thus allowing predictions for arbitrary choices of parameter values.

Values for all three parameters are assigned within the mould generation process for each simulation input. Values for $\Delta$ were generated from a uniform random deviate with fixed range $\Delta\in[0.05,0.95)$.  Both $\overline{\kappa}_{\rm 3N}$ and $\overline{\kappa}_{\rm 4N}$ were jointly generated from a 2D uniform distribution with $\overline{\kappa}_{\rm 3N}\in[0.1,0.71)$ and $\overline{\kappa}_{\rm 4N}\in[0.01,0.5)$, with an additional bound of $\overline{\kappa}_{\rm 4N} \leq \overline{\kappa}_{\rm 3N} / \sqrt{2}$. An additional set of 500 simulations (approximately 1/8th of the total) was created using instances where $\Delta$ was set to either $0.05$ or $0.95$, in order to improve the representation of cases at the edges of the parameter space.

CONDOR simulations were made to simulate growth of cultured tissues in the randomly created moulds for a set of randomly selected parameter values. A total of of $3569$ unique simulations were run in parallel on a multi-core machine and were divided into two disjoint sets to be used for training and evaluation of the machine learning model. Of these, $445$ were set aside to use as test data, while the remaining $3124$ were used in training and validation.

\subsection{Training Data Setup}
\label{sec:training_data_setup}

Results from CONDOR simulations initially consist of a set of cell positions, orientations and ECM bond tensions; effectively, a list of cells and their individual properties. We transform this data into a set of continuous 2D distributions to be plotted as 256$\times$256 pixel images that are compatible for training using the specific machine learning model we use (\texttt{pix2pix}). The value for a cell property in each pixel ($Z_{p}$) is calculated using a weighted sum from all nearby cells via,
\begin{equation}
\label{eqn:kde_pixel_sum}
Z_{p} = \sum_{i} z_{i} w_{i,p}
\end{equation}
where $z_{i}$ is the value an individual cell property, and $w_{i,p}$ is the corresponding weight. As in Ref. \cite{andrews2023}, upper case letters denote continuous field properties and lower case letters discrete cell properties. We use $z_{i}=1$ to calculate the density field (hence making density a relative measure of number of cells per pixel in the image). Weights are calculated using a multivariate Gaussian density function centered on a cell position and integrated over the pixel area,
\begin{equation}
\label{eqn:kde_normal_distribution}
w_{i,p} = \int_{x_{p}-\frac{l_{p}}{2}}^{x_{p}+\frac{l_{p}}{2}} \int_{y_{p}-\frac{l_{p}}{2}}^{y_{p}+\frac{l_{p}}{2}} \frac{e^{ -\left( \left( x-x_{i} \right)^{2} + \left( y-y_{i} \right)^{2} \right) /2\sigma^{2} }}{2 \pi \sigma^{2}} {\rm d}y {\rm d}x
\end{equation}
where $x_{i}$ and $y_{i}$ are the cell coordinates from the CONDOR simulation, $x_{p}$ and $y_{p}$ are the central coordinates of the pixel; these are all in units of $l_{p}$. $\sigma$ sets the width of the Gaussian; we set $\sigma = 3l_{p}$ to ensure the resulting image is smooth but not lacking structure. All property fields in the training data are uniformly scaled to range from $-1$ to $1$.

All continuous fields are inherently proportional to the density field. To ensure that the sum over pixels of the output fields agrees with averages over cell properties they are normalised via:
\begin{equation}
\label{eqn:norm_factor_density}
\overline{Z_{p}} = \frac{N_{p}}{\sum_{i p} w_{i,p}} \sum_{i} z_{i} w_{i,p} = \frac{N_{p}}{N_{c}} \sum_{i} z_{i} w_{i,p}
\end{equation}
where $N_{p}$ is the total number of pixels and $N_{c} = \sum_{ip} w_{i,p}$ is the total number of cells.

Augmentations were also subsequently applied to the training and validation set. These consisted of the same combinations of multiple transformations applied to both mould shapes (input images) and simulation results (output images), including mirroring in both the $x$ and $y$ axes and rotations in increments of $90$ degrees. Any duplicate moulds arising due to underlying mould symmetries were removed. The total number of training and validation examples including augmentations was $12653$. 

\subsection{Machine Learning}
\label{sec:ml}

The RAPTOR machine learning approach developed here is adapted from the \texttt{pix2pix} conditional generative adversarial network (cGAN), which we implemented using the TensorFlow framework using an almost identical cGAN to CONDOR-ML in Ref. \cite{andrews2023}. The key extension to this previous work is the inclusion of the model parameters $\Delta$ and $\overline{\kappa}$ as inputs to the machine learning approach; the model and neural network architecture itself is otherwise unchanged. This allows for the prediction of properties associated with a broader range of cultivated tissue types.

Since the machine learning framework is introduced in detail in Ref. \cite{andrews2023}, we focus on the differences here. The \texttt{pix2pix} model learns a mapping between input and output images. For our implementation 256$\times$256 pixel images were used for input and output. The input consisted of 5 channels representing the mould design and the simulation parameters. The output comprised 8 channels predicting the properties of both matrix and cells. The pixel size was set to $l_{p} = 125 l_{0} / 512$. \texttt{pix2pix} was used because it has a suitable architecture for image prediction, but has fewer weights than e.g. a stable diffusion model architecture. Since the network architecture differs slightly from that of CONDOR-ML due to these extra input layers, we have summarised the layers of the modified architecture in \ref{sec:appendixpix2pix}.

The five input layers are as follows: Two input layers describe the mould design, representing the layout of the mould depression (the initial shape of the cellular hydrogel) and the tether placement respectively. Both layers are arrays of boolean values representing the presence or absence of the depression or tethers respectively. Boolean values in the depression layer relating to tether locations are set to false. Three additional input layers correspond to the parameters used in the simulation; $\Delta$, $\overline{\kappa}_{\rm 3N}$ and $\overline{\kappa}_{\rm 4N}$. They are input into the model as 256$\times$256 arrays with a single repeated value of the parameter for compatibility with the pix2pix architecture (in other words, additional image channels).

The eight output layers of the model correspond to the properties of the cultured tissue predicted by the simulation: the cell density ($\rho_{i}$), orientation vector products ($s_{x,i}^{2}$, $s_{y,i}^{2}$, $s_{z,i}^{2}$, $s_{x,i} s_{y,i}$, $s_{x,i} s_{z,i}$ and $s_{y,i} s_{z,i}$), and average strain in adjacent bonds ($\tau_{i}=\tau_{0,i}/\kappa_0 l_{0}$, where $\tau_{0,i}$ is the average tension in adjacent bonds).

The model was trained on a laptop-grade GPU (M1 MacBook Pro), with a total training time of approximately 12 hours.

Comparison between the shapes of CONDOR and RAPTOR results was made by calculating the Jaccard distance. For practical purposes, this is the ratio of the number of pixels that have the same value after applying a density threshold (0.2 cells / pixel) to the total number of pixels in the images. The same threshold was used to calculate area and width.

\subsection{Tissue growth}

Corneal, fibroblast, muscle and tendon tissue growth results have been previously reported (see Refs. \cite{garvin2003a,capel2019a, mukhey2018,kostyuk2004}), and protocols for growth of those tissues can be found in those papers. We briefly summarise them below.

Human corneal stromal stem cell data came from Ref. \cite{mukhey2018}, and for full details see the information published there. We provide a brief summary: The hydrogel was formed of type I rat tail collagen (80\%), 10x Minimum Essential Medium from Invitrogen,
Paisley, UK (10\%) and Neutralizing Solution from TAP Biosystems, Royston, Hertfordshire, UK (5.9\%). Cells were seeded in the hydrogel at a density of $6.7\times 10^{5}$ cells/mL and cultured for 8 hours. H-shaped (I shaped) moulds containing tethering bars were manufactured from PEEK. Contraction was reported in Fig. 2(b), and we read contraction information from that figure.

Fibroblast data came from Ref. 
\cite{kostyuk2004}, and for full details see the information published there. We again provide a brief summary: The hydrogel was formed from 2.03mg/ml type I rat tail collagen, which was mixed with 12.5\% 10x Dulbecco's modified Eagle's medium, with the resulting solution neutralised with 5M NoOH. Cells were seeded in the hydrogel mixture at at a density of $1\times 10^{6}$ cells/ml. 5ml of the cell-laden hydrogel mixture was added to a mould consisting of a rectangular shape (56 mm $\times$ 24 mm $\times$ 3.7 mm) with plastic tethering bars located at each end of the long axis. We determined the area contraction information from the image in Fig. 1(b) of Ref. \cite{kostyuk2004}. Width contraction is reported in Fig. 7 of Ref. \cite{kostyuk2004} as a geometric factor.

Muscle tissue growth was reported in Ref. \cite{capel2019a}. Capel et al \cite{capel2019a} used C$_2$C$_{12}$ myoblasts in 85\% rat tail collagen type I hydrogel with 10\% (v/v) MEM neutralised with NaOH. They seeded cells at $4\times 10^{6}$ cells/ml in 5\% growth medium. Growth medium was formed from Dulbecco’s Modified Eagle’s Medium (DMEM,
Fisher–Scientific, UK), 20\% fetal bovine serum (FBS, Pan Biotech,
UK), and, 1\% Penicillin/Steptomyocin (P/S, Fisher–Scientific,
UK). They printed moulds from (1) polylactic acid (PLA) using fused deposition modelling (FDM) and (2) polyamide-12 using laser sintering (LS).

Tendon tissue growth was reported in Ref. \cite{garvin2003a}, and for full details see the information provided there. In brief: Garvin et al \cite{garvin2003a} used avian tendon internal fibroblasts suspended at a density of 200,000 per 170$\mu$l of Type I collagen (Vitrogen; Cohesion Technologies, Palo Alto, CA). They were then added to a Tissue Train culture plate (Flex-cell International, Hillsborough, NC) before incubation for up to 14 days.

Glial tissue growth was previously reported in thesis form (Ref. \cite{orourke2015}). C6 glial cells were used, and seeded at $4 \times 10^{6}$ cells/ml in a rat tail type I collagen hydrogel and incubated for 24 hours. Moulds incorporated tethering posts, and were manufactured from PEEK by TAP Biosystems (polyetheretherketone) using CNC machining. Collagen hydrogels were formed by neutralising an acidic solution of 2 mg/ml Type 1 collagen. By slowing the gelation process through keeping the reagents cold (4$^\circ$ C) there is time to mix the neutralised collagen solution with the cell suspension. The mixture is then cast into a mould and warmed to 37$^\circ$ C where it sets to form a hydrogel in 10-30 minutes. 

Comparisons of tissue contraction were made using the width and area. These were determined by tracing around tissue in images from Refs. \cite{garvin2003a,capel2019a, mukhey2018,kostyuk2004} using InkScape \cite{inkscape}.

\subsection{Workflow}

The process of training and validation involved the following steps:
\begin{enumerate}
\item Generate a large set of random mould designs.
\item Run CONDOR simulations for the large set of designs to create a training data set.
\item Convert the output of the CONDOR simulations into images for training RAPTOR.
\item Split the data set into training and validation groups for machine learning.
\item Train and validate the RAPTOR machine learning using this data set.
\item Fit RAPTOR and CONDOR outputs against experimental data to find model parameters (i.e. $\Delta$ and $\kappa$) that lead to the same width and area contractions found in the experimental data.
\item Run a final additional CONDOR simulation with the same parameters as RAPTOR to check RAPTOR outputs.
\end{enumerate}

\section{Results}
\label{sec:results}

\subsection{Predictive capability of RAPTOR}
\label{sec:raptortest}

\begin{figure*}[hp]
\centering
{
\includegraphics[width=0.8\textwidth]{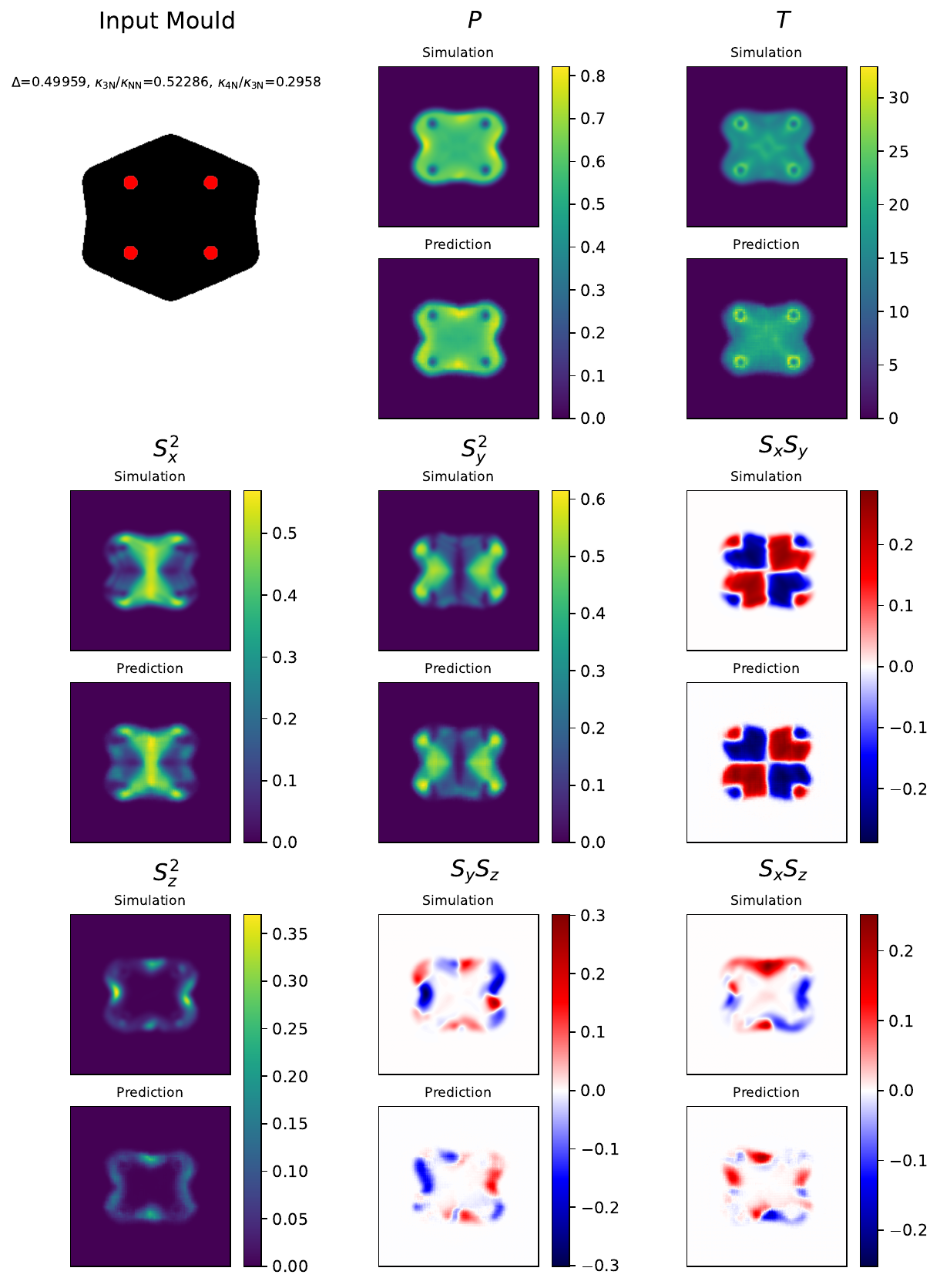}
\caption{Comparison between the results of the CONDOR simulation and the machine learning prediction for an example mould from our test data set, shown in the top left panel. The remaining panels show comparisons between all 8 target fields produced by the CONDOR simulation and predictions made by RAPTOR. This includes dimensionless cell density $P$, dimensionless tension $T$, and orientation products ($Q$ tensor) $S_{x}^2$, $S_{y}^2$, $S_{z}^2$, $S_{x} S_{y}$, $S_{y} S_{z}$ and $S_{x} S_{z}$. All properties are dimensionless, and comparisons between individual properties are shown on the same colour scale.}
\label{fig:mould_1_results_comparisons}
}
\end{figure*} 

\begin{figure*}[hp]
\centering
\includegraphics[width=0.85\textwidth]{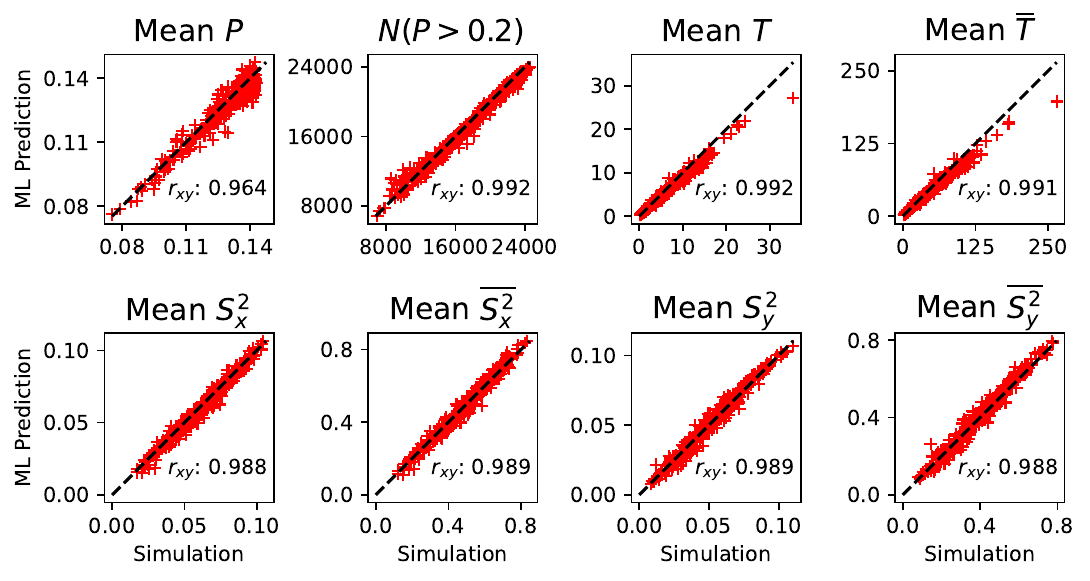}
\caption{Figure showing bulk comparisons for the entire test set (which has a sample size of 445 cases). Properties include mean pixel density $P$, number of pixels with density exceeding $0.2$ as a proxy for area, $N \left( P > 0.2 \right)$, and mean values for both initial and normalised tension ($T$ and $\overline{T}$), $x$ alignment ($S_{x}^{2}$ and $\overline{S_{x}^{2}}$) and $y$ alignment ($S_{y}^{2}$ and $\overline{S_{y}^{2}}$). Also shown for each comparison is the Pearson correlation coefficient, $r_{xy}$, and a dashed diagonal to indicate the line of perfect agreement. All properties are dimensionless. Barred and unbarred properties are related by a scaling factor depending on $N_{c}$. Note that the the axes of the two plots for $P$ and $N(P>0.2)$ do not start at 0.}
\label{fig:statistics_comparisons_1}
\end{figure*}

\begin{figure*}[hp]
\centering
\includegraphics[width=0.85\textwidth]{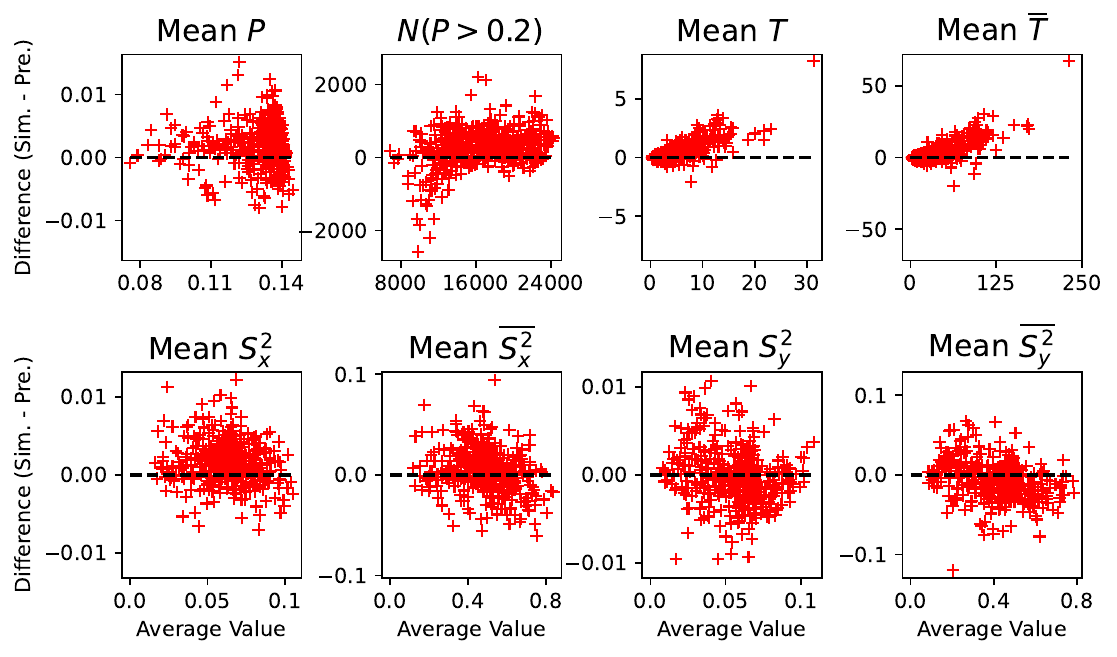}
\caption{Figure showing Bland-Altman comparisons for the entire test set (which has a sample size of 445 cases), corresponding to the data shown in Figure \ref{fig:statistics_comparisons_1}. Properties include mean pixel density $P$, number of pixels with density exceeding $0.2$ as a proxy for area, $N \left( P > 0.2 \right)$, and mean values for both initial and normalised tension ($T$ and $\overline{T}$), $x$ alignment ($S_{x}^{2}$ and $\overline{S_{x}^{2}}$) and $y$ alignment ($S_{y}^{2}$ and $\overline{S_{y}^{2}}$). We note that the Bland-Altman plot shows a systematic underestimation of higher tension by RAPTOR.}
\label{fig:statistics_comparisons_bland_altman_1}
\end{figure*}

The predictive capacity of the trained RAPTOR model is initially tested against the test data set of $445$ CONDOR simulations that was set aside from the training data (see Sec. \ref{sec:trainingdata}).

A representative comparison of tissue properties shown in Fig.~\ref{fig:mould_1_results_comparisons} demonstrates good qualitative agreement between CONDOR simulation results and RAPTOR predictions for an example mould in the test set. For this case the mould has double mirror symmetry and intermediate valued simulation parameters ($\Delta = 0.49959$, $\kappa_{\rm 3N}/\kappa_{\rm NN}=0.52286$ and $\kappa_{\rm 4N}/\kappa_{\rm 3N}=0.29583$). Most of the output fields show good qualitative agreement, including density $P$, tension (strain) $T$, and alignment product fields $S_{x}^{2}$, $S_{y}^{2}$ and $S_{x} S_{y}$. The density field shows accurate prediction of the shape of the cultured tissue. The strain (tension) has approximately the correct range and structure, although some of the finer structure is missing. Predictions of $S_{x}S_{y}$ are accurate and the correct positions for sign changes are predicted. $S_{x}^{2}$ and $S_{y}^{2}$ predictions are very accurate, and the model has additionally learned that these fields are approximate inverses of each other. The fields depending on $S_{z}$ (which are less important since cells are mostly directed within the $xy$-plane), $S_{z}^{2}$, $S_{y} S_{z}$ and $S_{x} S_{z}$, show good magnitude, but the locations of the sign changes are shifted.

Quantification of the agreement can be found from comparisons between the mean values of tissue properties for RAPTOR and CONDOR predictions show good agreement across the entire test data set. Figure~\ref{fig:statistics_comparisons_1} shows comparisons of mean density $P$, total tissue area (represented by the number of pixels with density $P>0.2$ cells/pixel, $N\left(P<0.2\right)$), and the mean values for tension $T$, $S_{x}^{2}$ and $S_{y}^{2}$ as well as their density normalised equivalents ($\overline{T}$, $\overline{S_{x}^{2}}$ and $\overline{S_{y}^{2}}$). Perfect agreement is indicated by the dashed diagonal line.

The scatter graphs show very good agreement, with points situated close to the line of perfect agreement and high Pearson correlation coefficient ($r_{xy}$) values. A single outlier can be seen in the comparisons for tension (both $T$ and $\overline{T}$) in which the machine learning model under-predicts an otherwise large mean tension value. This corresponds to a culture simulated with a very high $\Delta = 0.931$.

We additionally use Bland-Altman graphs to compare deviation between predicted and simulated outputs, as shown in Fig.~\ref{fig:statistics_comparisons_bland_altman_1}. These also indicate good agreement across all value measures, with offsets typically no more than 10\% of the total range of values. The exception to this is the  previously noted outlier in tension comparisons. This is shown by the Bland-Altman plot to be part of a systematic underestimation of tension by RAPTOR. Higher tension cases are typically high contraction and we think this underestimate relates to underestimates of contraction by RAPTOR for high contraction (see limitations).

\subsection{Model parameters and predictive capability}
\label{sec:raptorresults}

The aim of this section is to compare CONDOR and RAPTOR results across the parameter space of $\Delta$, $\kappa_{\rm 3N}$ and $\kappa_{\rm 4N}$. Two standard mould designs were used for this analysis: a rectangular test mould design with a 1:2 height to width ratio with tethers placed in the four corners of the mould, and an I-shaped mould with a broad tethering bar without an analogue in the training data set. The 4 pillar model is equivalent to the mould analysed in Refs. \cite{orourke2017} and \cite{hague2019a}.

\begin{figure*}[hp]
\centering
\includegraphics[width=0.6\textwidth]{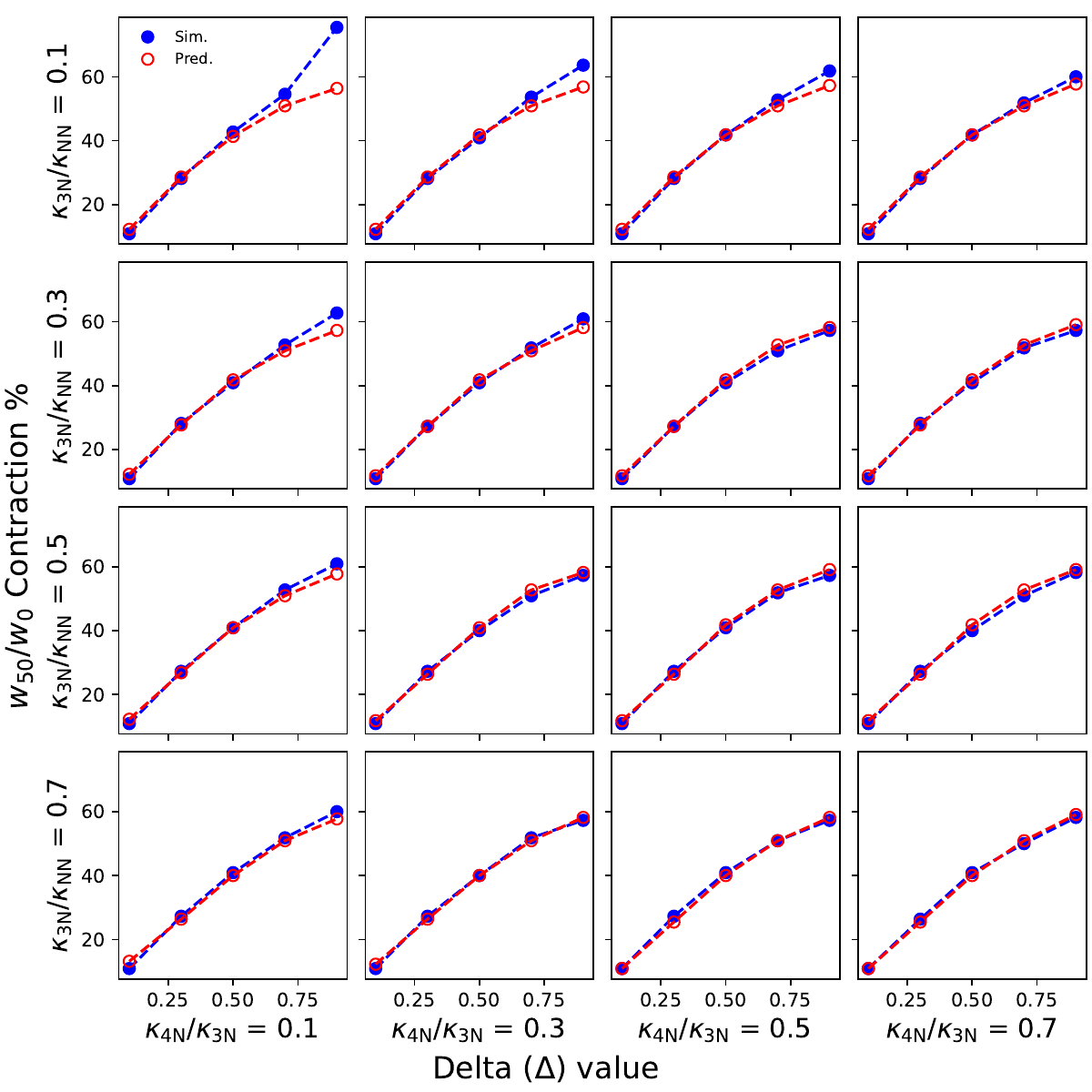}\\[10pt]
\includegraphics[width=0.6\textwidth]{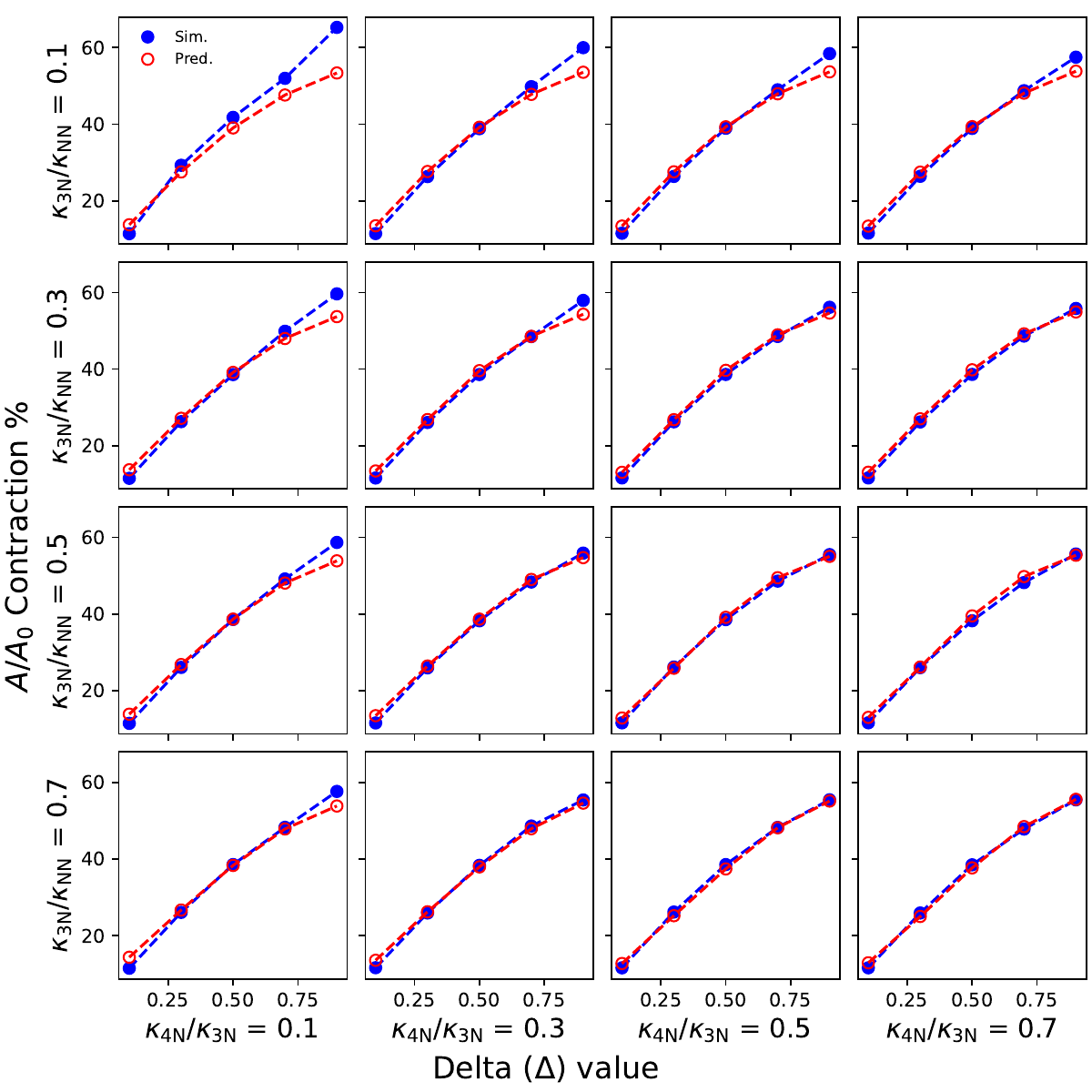}
\caption{Predicted and simulated contractions for a four pin rectangle mould design (see Ref. \cite{orourke2015,hague2019a}) for multiple combinations of $\kappa_{\rm 3N}/\kappa_{\rm NN}$ and $\kappa_{\rm 4N}/\kappa_{\rm 3N}$ values. Each point shows measurements of percentage decrease in central width of the cell matrix in relation to simulated $\delta$ value. Agreement is excellent across the majority of the parameter space, with the largest differences for large $\Delta$ and small $\kappa$.}
\label{fig:contraction_curves_dogbone}
\end{figure*}

\begin{figure*}[hp]
\centering
\includegraphics[width=0.6\textwidth]{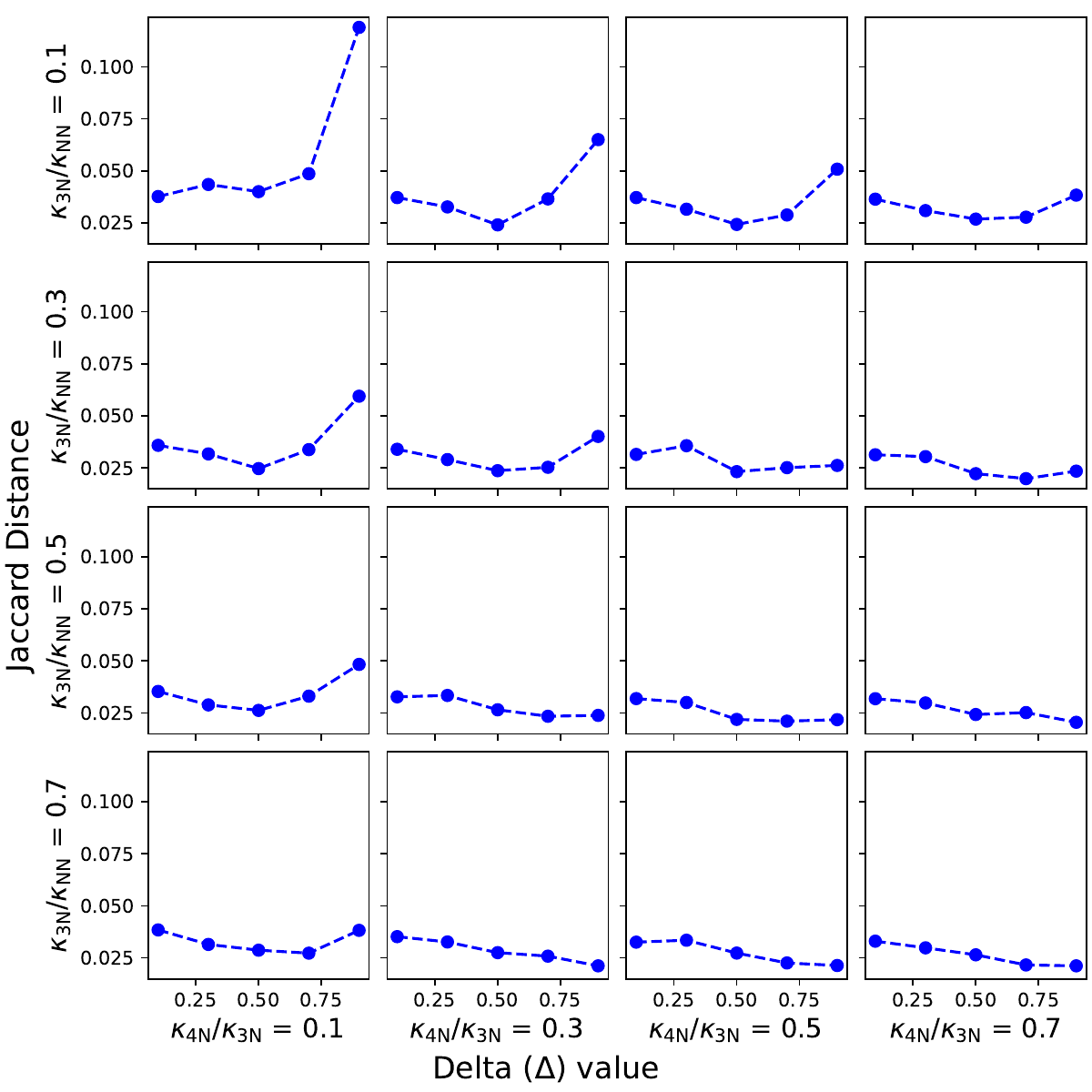}\\[10pt]
\caption{Jaccard distance between predicted and simulated contractions for a four pin rectangle mould design (see Ref. \cite{orourke2015,hague2019a}) for multiple combinations of $\kappa_{\rm 3N}/\kappa_{\rm NN}$ and $\kappa_{\rm 4N}/\kappa_{\rm 3N}$ values. There is less than 5\% difference between tissue shapes over the majority of the parameter space, but the difference increases for large $\Delta$ and small $\kappa$ when simulated tissue is very contractile.}
\label{fig:contraction_curves_jaccard_dogbone}
\end{figure*}

Figure~\ref{fig:contraction_curves_dogbone} shows contractions obtained from CONDOR simulations and RAPTOR predictions for the rectangle mould design for various $\kappa_{\rm 3N}/\kappa_{\rm NN}$ and $\kappa_{\rm 4N}/\kappa_{\rm 3N}$ values. CONDOR analysis for glial tissue grown within this type of mould can be found in Ref. \cite{hague2019a}. The upper panels show the width contraction (the percentage decrease of the predicted tissue width as compared to the width of the mould) as a function of $\Delta$. The lower panels show the percentage area contraction. Figure \ref{fig:contraction_curves_jaccard_dogbone} shows the Jaccard distance between CONDOR and RAPTOR results for this design. 

Overall there is close agreement between both width and area contraction predicted by CONDOR and RAPTOR. The contraction differs primarily for cases with low $\kappa_{\rm 3N}/\kappa_{\rm NN}$, low $\kappa_{\rm 4N}/\kappa_{\rm 3N}$ and high $\Delta$, with the largest difference seen for $\Delta = 0.9$, $\kappa_{\rm 3N} = 0.1$ and $\kappa_{\rm 4N}/\kappa_{\rm 3N} = 0.1$. Quantitative agreement in shape measured by the Jaccard distance is excellent, with a less than 5$\%$ difference over the majority of the parameter space. As indicated by the contraction, agreement decreases for large $\Delta$ and small $\kappa$ when simulated tissue is very contractile, with the difference rising to around 12\% in that case. This difference may show up as the underestimate of tension by RAPTOR in Fig. \ref{fig:statistics_comparisons_1}, since increased contraction is associated with higher tension.

\begin{figure*}[hp]
\centering
\begin{subfigure}{\textwidth}
\subcaption{CONDOR}
\includegraphics[width=0.82\textwidth]{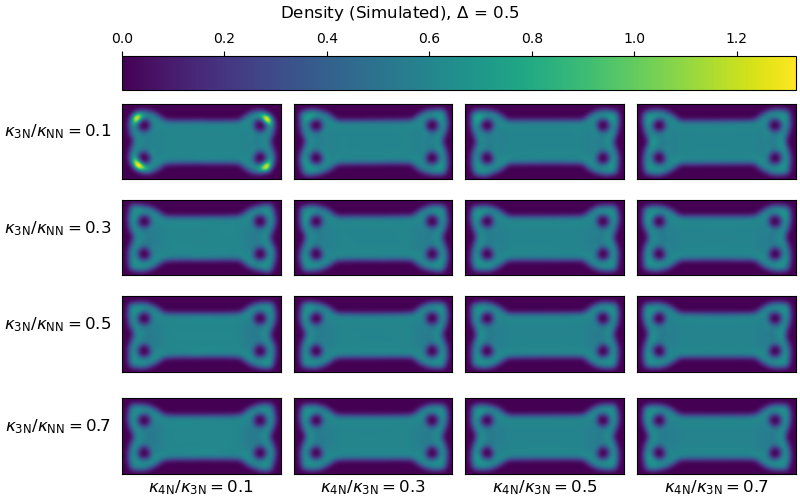}
\end{subfigure}
\hspace{20pt}

\begin{subfigure}{\textwidth}
\subcaption{RAPTOR}
\includegraphics[width=0.82\textwidth]{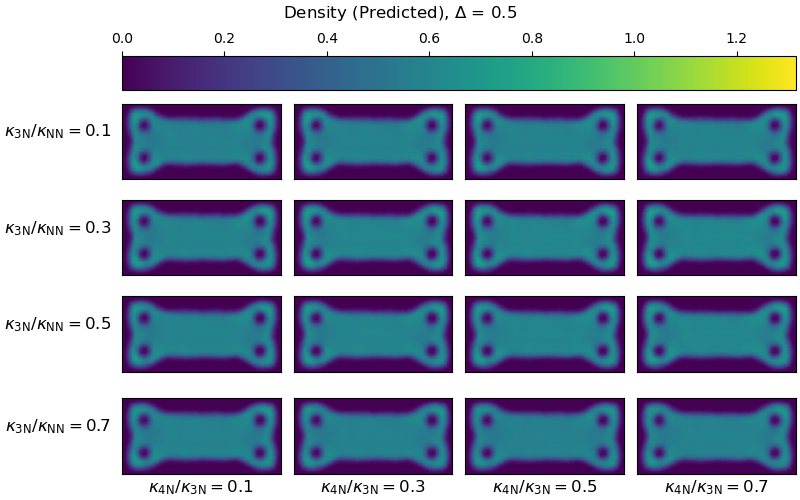}
\end{subfigure}
\caption{Figure showing the density distributions for a four pin rectangle mould design (see Ref. \cite{hague2019a}) simulated using CONDOR and predicted using RAPTOR with $\Delta = 0.5$ and varying $\kappa_{\rm 3N}/\kappa_{\rm NN}$ and $\kappa_{\rm 4N}/\kappa_{\rm 3N}$ values.}
\label{fig:varying_density_dogbone_simulation_d0x5}
\label{fig:varying_density_dogbone_prediction_d0x5}
\end{figure*}

There is a very good qualitative agreement between tissue shapes predicted for small to medium sized $\Delta$ values across the parameter space of $\kappa$. As an example, Fig.~\ref{fig:varying_density_dogbone_simulation_d0x5} shows comparison between tissue shapes for the 4 pin rectangular design with $\Delta=0.5$. The top panels show CONDOR simulations and the bottom panels RAPTOR predictions. RAPTOR predicts a slightly lower density than CONDOR on the outside edge the tethering pillars for $\kappa_{\rm 3N}/\kappa_{\rm NN}=0.1$ and $\kappa_{\rm 4N}/\kappa_{\rm 3N}=0.1$. Qualitative agreement between the shapes is otherwise good. 

\begin{figure*}[hp]
\centering
\begin{subfigure}{\textwidth}
\subcaption{CONDOR}
\includegraphics[width=0.82\textwidth]{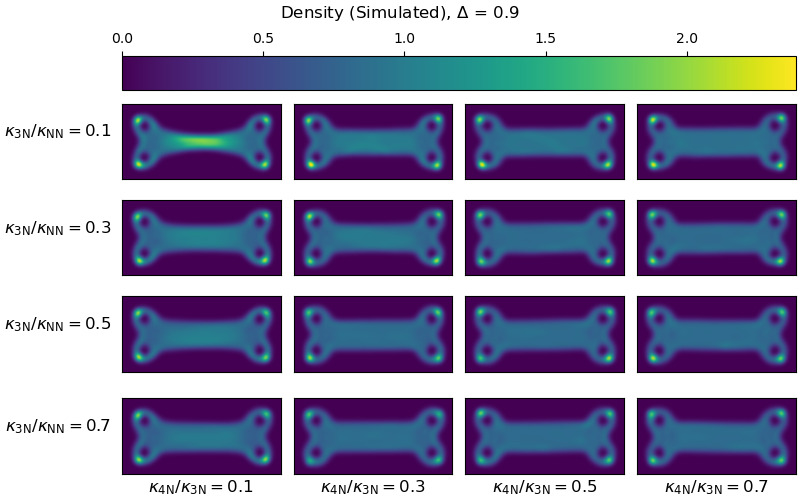}
\end{subfigure}
\hspace{20pt}

\begin{subfigure}{\textwidth}
\subcaption{RAPTOR}
\includegraphics[width=0.82\textwidth]{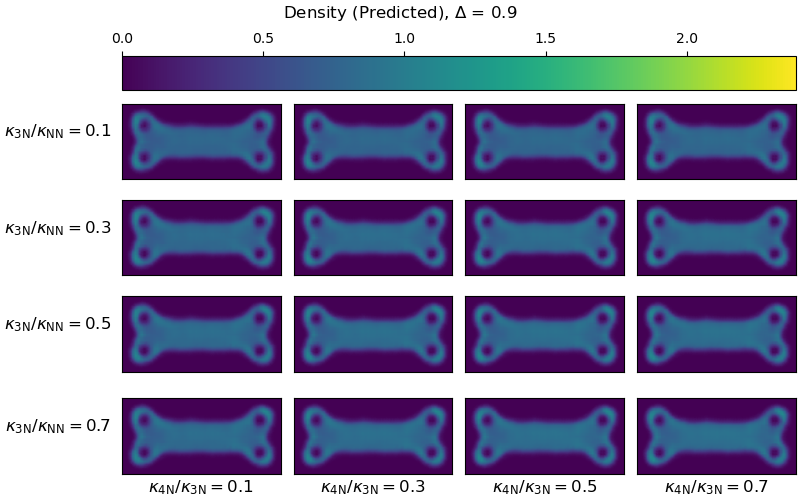}
\end{subfigure}
\caption{Figure showing the predicted density distributions for a four pin rectangle mould design (see Ref. \cite{hague2019a}) determined using CONDOR and predicted using RAPTOR with $\Delta = 0.9$ and varying $\kappa_{\rm 3N}/\kappa_{\rm NN}$ and $\kappa_{\rm 4N}/\kappa_{\rm 3N}$ values.}
\label{fig:varying_density_dogbone_simulation_d0x9}
\label{fig:varying_density_dogbone_prediction_d0x9}
\end{figure*}

The origin of the differences between CONDOR and RAPTOR at the highest contractions can be established from direct comparison of tissue shapes for $\Delta=0.9$ shown in Fig.~\ref{fig:varying_density_dogbone_simulation_d0x9}. For $\kappa_{\rm 3N}/\kappa_{\rm NN} = 0.1$ and $\kappa_{\rm 4N}/\kappa_{\rm 3N} = 0.1$, there is an increased curvature and corresponding increase in tissue density close to the centre of the mould as the cell-induced contraction overcomes the stiffness of the bonds. Away from high $\Delta$, and low $\kappa$, agreement is very good between both simulated and predicted results across the whole parameter space.

\subsubsection{Predictive capability for structures with no analogue in the training data}

\begin{figure*}[hp]
\centering
\includegraphics[width=0.6\textwidth]{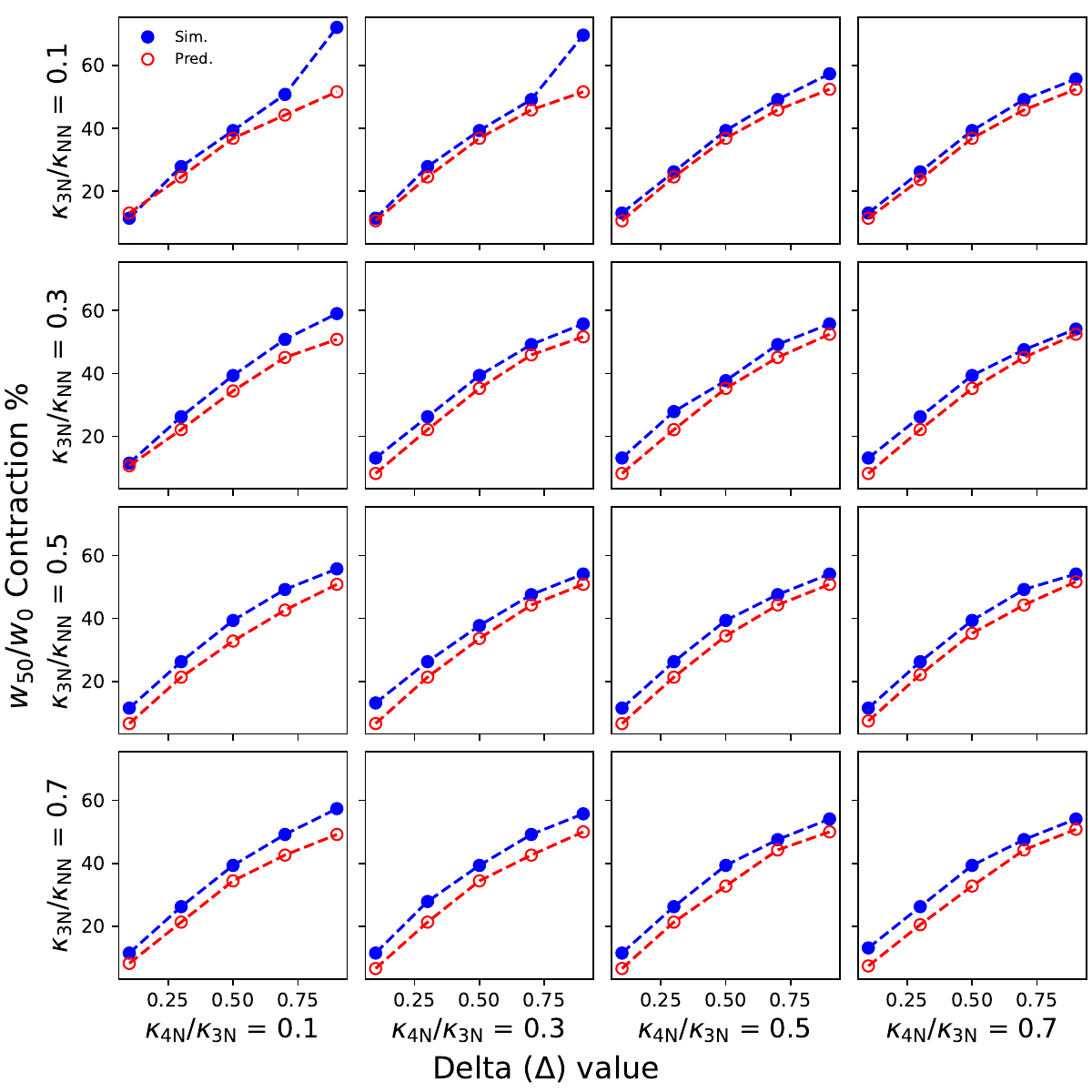}

\includegraphics[width=0.6\textwidth]{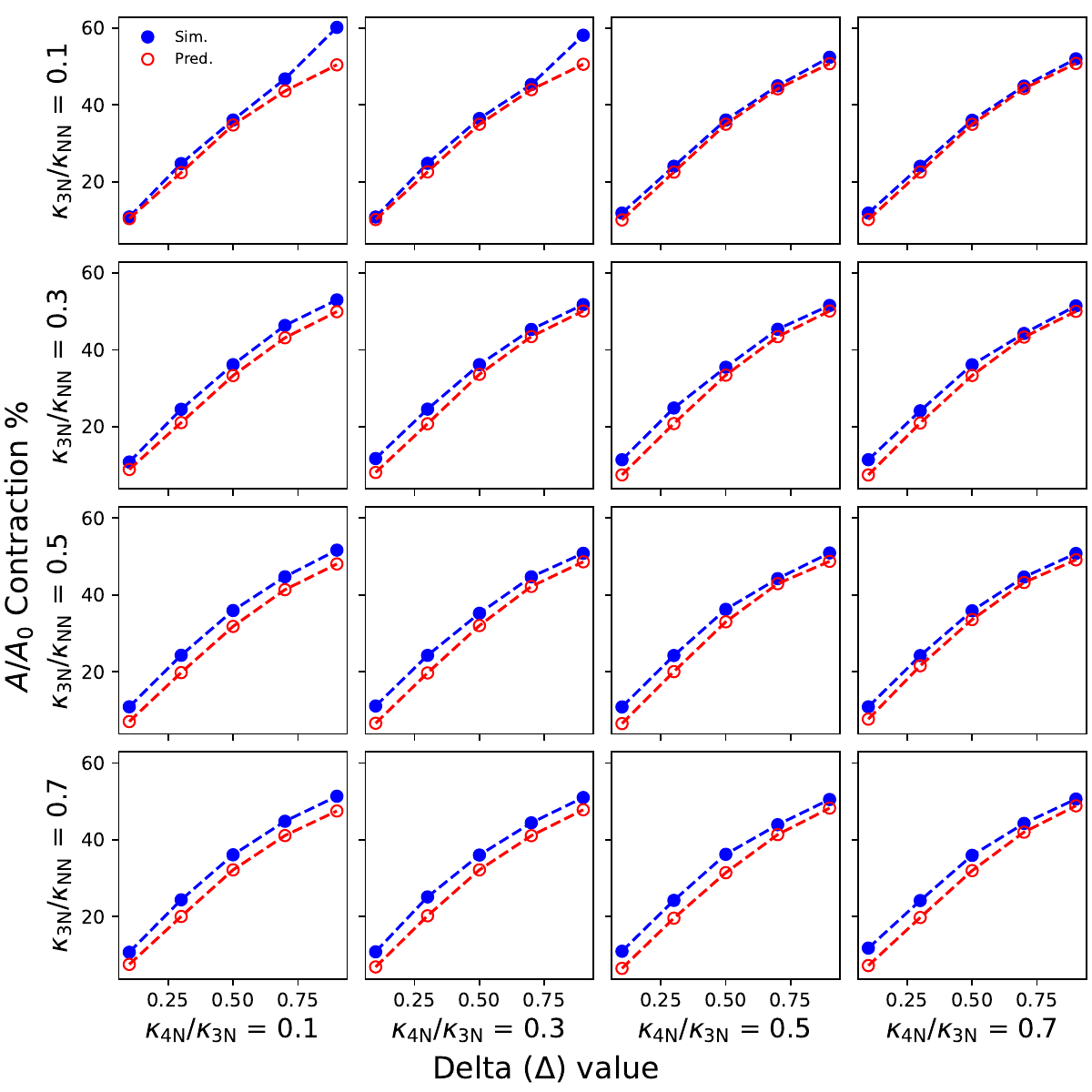}
\caption{Glial tissue, mould from Ref. \cite{east2010} / Fig. \ref{fig:real_comparison_mukhey} , corneal tissue, mould from Ref. \cite{mukhey2018}. Figure showing the $w_{50}/w_0$ and $A/A_{0}$ contraction curves for a 2 pin rectangle mould design for multiple combinations of $\kappa_{\rm 3N}/\kappa_{\rm NN}$ and $\kappa_{\rm 4N}/\kappa_{\rm 3N}$ values. Each contraction curve shows measurements of percentage decrease in central width of the cell matrix in relation to simulated $\delta$ value. Mould was experimentally used to grow glial (see Ref. \cite{east2010}) corneal tissue (see Ref. \cite{mukhey2018}).}
\label{fig:contraction_curves_east}
\end{figure*} 

\begin{figure*}[hp]
\centering
\includegraphics[width=0.6\textwidth]{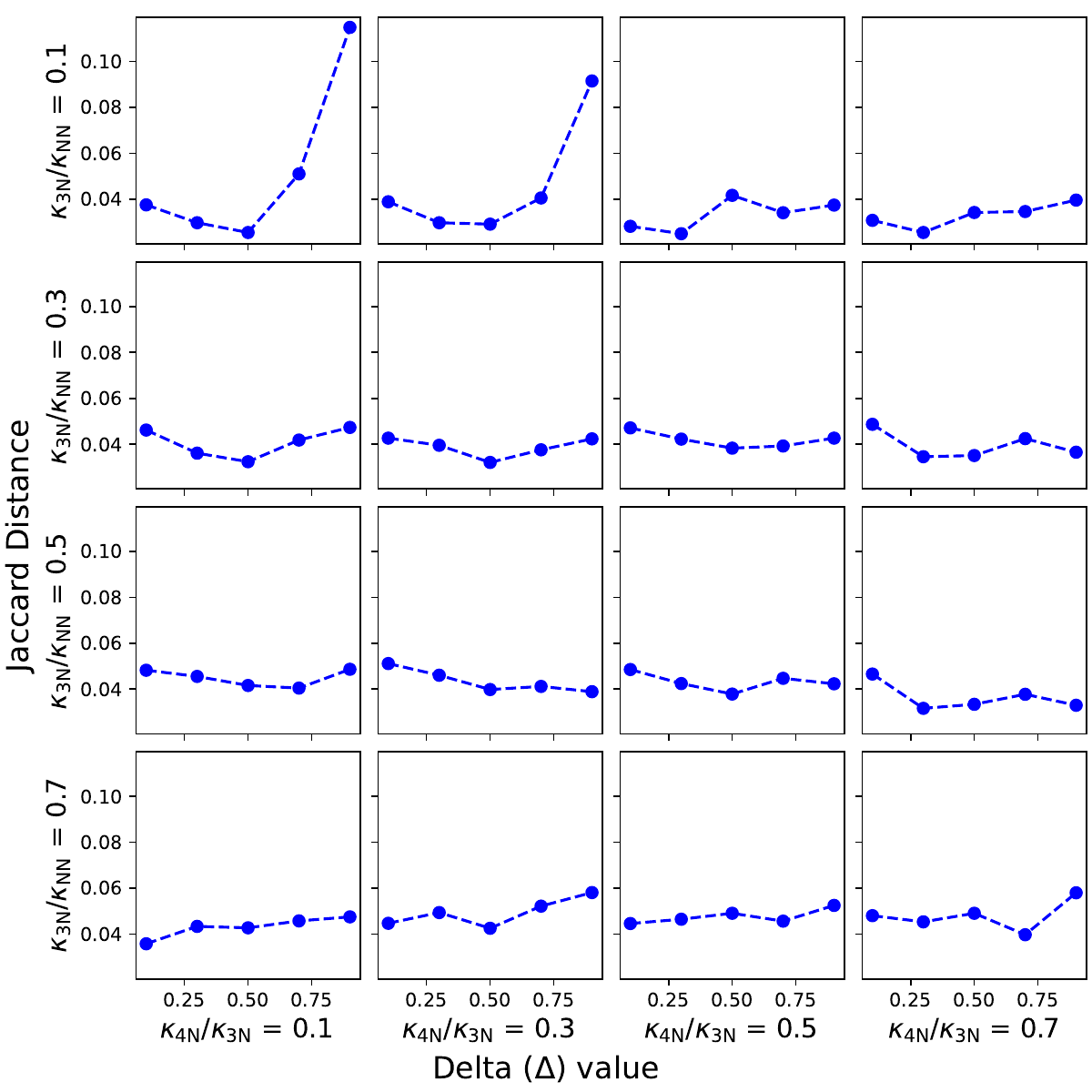}
\caption{Glial tissue, mould from Ref. \cite{east2010} / Fig. \ref{fig:real_comparison_mukhey} , corneal tissue, mould from Ref. \cite{mukhey2018}. Figure showing the Jaccard distance for a 2 pin rectangle mould design for multiple combinations of $\kappa_{\rm 3N}$ and $\kappa_{\rm 4N}/\kappa_{\rm 3N}$ values. Agreement is good across the majority of the parameter space, but with larger differences for small $\kappa$ and large $\Delta$ where contraction is large.}
\label{fig:contraction_curves_jaccard_east}
\end{figure*}

The I-shaped mould examined in this section contains a large continuous tethering bar, which is used to show how RAPTOR handles unexpected tethering arrangements. The training data set contains random arrangements of many small circular tethers, so the case of a long continuous tethering bar has no analogue within the training data set. Figure \ref{fig:contraction_curves_east} shows comparisons between contractions from CONDOR simulations and RAPTOR predictions for an I-shaped mould with a long tethering bar and similar dimensions to those used to grow both glial (Ref. \cite{east2010}) and corneal tissue cultures (Ref. \cite{mukhey2018}). The width contraction is underestimated by RAPTOR, with differences becoming worse for large $\Delta$ and small $\kappa$. Similar area contractions are predicted by both RAPTOR and CONDOR except for the specific case with $\Delta=0.9$ and small $\kappa$. 

Fig. \ref{fig:contraction_curves_jaccard_east} shows the Jaccard distance for this mould to provide a quantitative measure of similarity. Agreement is good across the majority of the parameter space (again less than 5\% difference), but with larger differences (again around 12\%) for small $\kappa$ and large $\Delta$ where contraction is large.

\begin{figure*}[hp]
\centering
\begin{subfigure}{\textwidth}
\subcaption{CONDOR}
\includegraphics[width=0.7\textwidth]{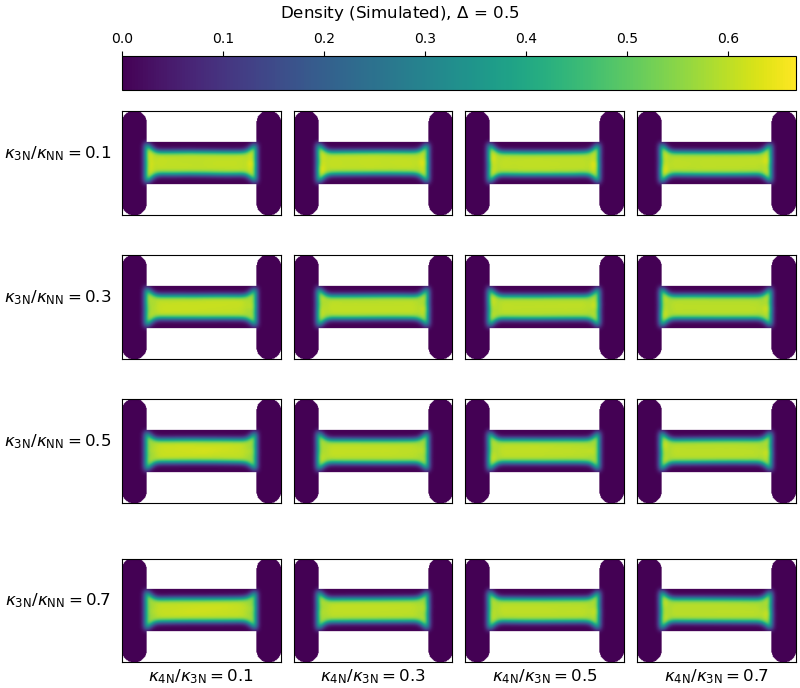}\\[10pt]
\end{subfigure}
\hspace{15pt}

\begin{subfigure}{\textwidth}
\subcaption{RAPTOR}
\includegraphics[width=0.7\textwidth]{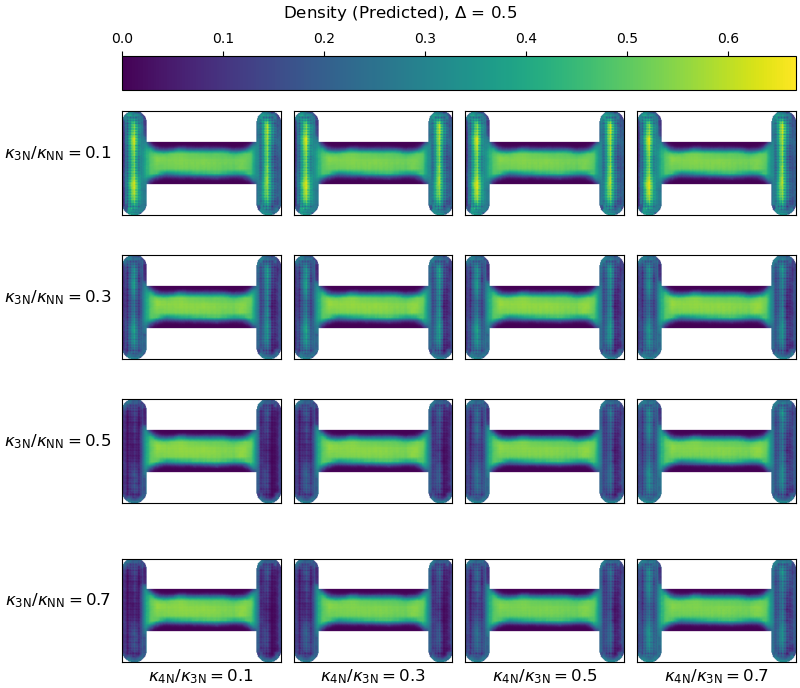}
\end{subfigure}
\caption{Density distributions for an I-shaped mould (used for Glial tissue in  Ref. \cite{east2010} and corneal tissue in Ref. \cite{mukhey2018}) predicted using CONDOR with $\Delta = 0.5$ and various $\kappa_{\rm 3N}/\kappa_{\rm NN}$ and $\kappa_{\rm 4N}/\kappa_{\rm 3N}$ values.}
\label{fig:varying_density_east_prediction_d0x5}
\end{figure*}

\begin{figure*}[hp]
\centering
\begin{subfigure}{\textwidth}
\subcaption{CONDOR}
\includegraphics[width=0.7\textwidth]{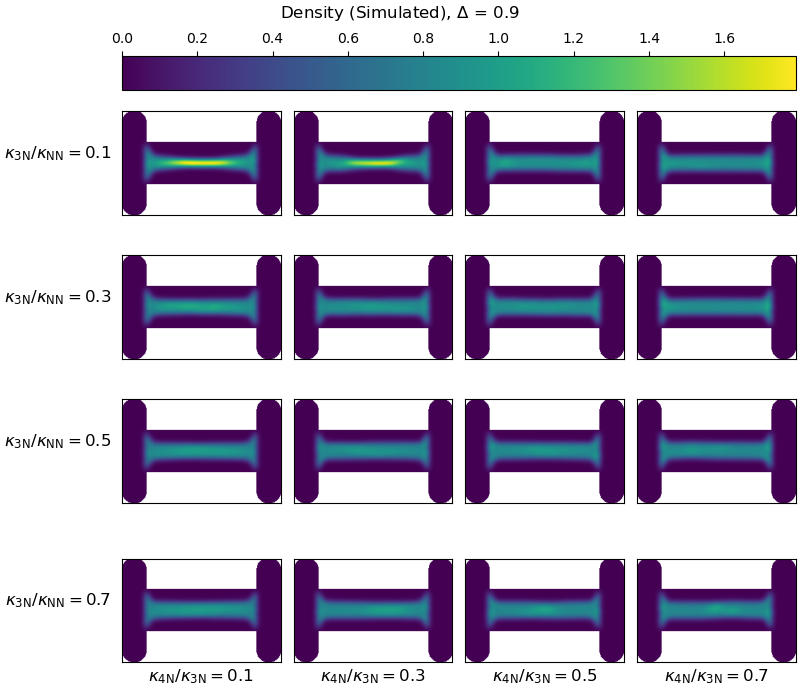}
\end{subfigure}
\hspace{15pt}

\begin{subfigure}{\textwidth}
\subcaption{RAPTOR}
\includegraphics[width=0.7\textwidth]{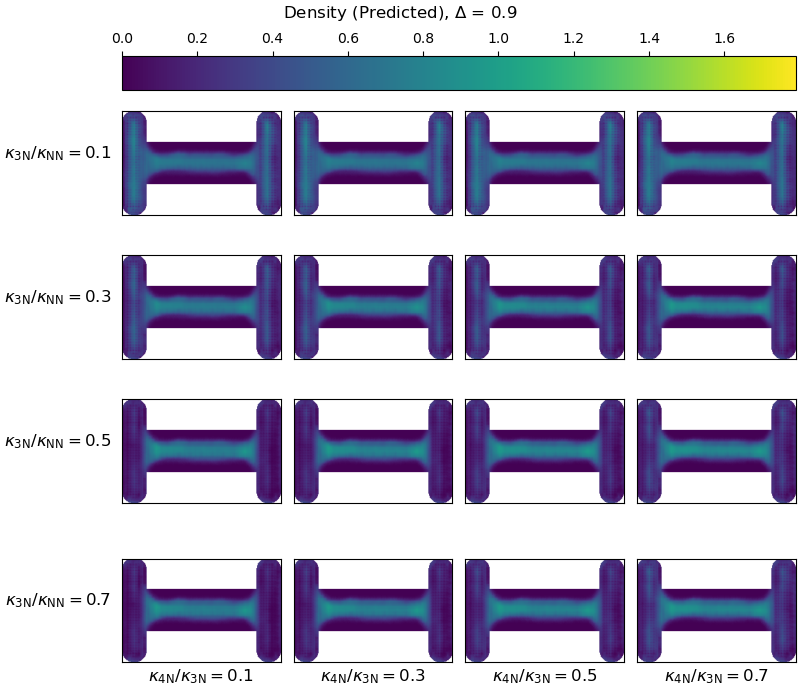}
\end{subfigure}
\caption{Density distributions for an I-shaped mould (used for Glial tissue growth in  Ref. \cite{east2010} and corneal tissue in Ref. \cite{mukhey2018}) predicted using CONDOR with $\Delta = 0.9$ and various $\kappa_{\rm 3N}/\kappa_{\rm NN}$ and $\kappa_{\rm 4N}/\kappa_{\rm 3N}$ values.}
\label{fig:varying_density_east_prediction_d0x9}
\end{figure*}

Figures \ref{fig:varying_density_east_prediction_d0x5} and \ref{fig:varying_density_east_prediction_d0x9} show density comparisons between CONDOR and RAPTOR predictions for various $\kappa$, with $\Delta=0.5$ and $\Delta=0.9$ respectively. Although spurious cell density can be seen within the long tethers (especially for $\Delta=0.5$ and $\kappa_{\rm 3N}=0.1$), the tissue shape within the depression matches the CONDOR simulations well across most of the parameter space (as measured quantitatively in Fig. \ref{fig:contraction_curves_jaccard_east} by the Jaccard distance). Again, it can be seen that within CONDOR simulations the cell induced contraction overcomes the bond stiffness for large $\Delta$ and small $\kappa$, and this is not well reproduced by the RAPTOR predictions probably due in part to the limited number of cases with this feature in the training data set. Given that there are no cases of extended tethering bars in the training data set (all cases involved circular tethers) it can be seen that the machine learning algorithm is also capable of extrapolation in some cases. (We note, however, that the model would be unlikely, for example, to be able to reliably extrapolate for tissues with very different $\kappa$ or $\Delta$ values). In practical terms, the tissue within tether bars is not used. In some cases it is not even seeded with cells (see Ref. \cite{lotfi2024}). So a spurious result within the tether bars is unimportant in practice.

In summary, the model has successfully learned the most general CONDOR properties, but was unable to learn behaviour in the small region of the parameter space where the contraction overcomes bond stiffness. For both mould types discussed in this section, most predictions from the machine learning algorithm are in good agreement with CONDOR regardless of $\Delta$, $\kappa_{\rm 3N}/\kappa_{\rm NN}$ and $\kappa_{\rm 4N}$ values. 

\subsection{Validation against experimentally cultured tissues and rapid determination of model parameters}
\label{sec:validation}

The primary purpose of this section is to validate the RAPTOR and CONDOR methods against experimental results for a variety of artificial tissue types grown in various tethered mould designs. The central width contraction and total area contraction were obtained for several experimental cases. We either (1) used images of the contracted 3D cultures for each mould to calculate (a) the ratio of tissue to mould width, and / or (b) the ratio of tissue to mould area or (2) used values reported in the literature. These ratios were then determined from both CONDOR simulations and RAPTOR predictions for direct comparison with the experiments.

The secondary purpose is to show how the speed of the machine learning algorithm can be leveraged to determine CONDOR model parameters using a fitting process. The RAPTOR method was used in a non-linear least squares fit to determine the values of $\Delta, \kappa_{\rm 3N}$ and $\kappa_{\rm 4N}$ that most closely match the tissue area and width of the experiments.

We use two separate methods of identifying optimal values for parameters. The first involves determination of parameter space values by non-linear least squares fitting by predicting tissue shapes with RAPTOR within the fit function; this is made possible by the speed of the RAPTOR algorithm. During the fit, the residual sum of squares (RSS) is minimised:
\begin{equation}
{\rm RSS} = \left( \bar{A} - A \right)^{2}/A_{0}^2 + \left( \bar{w}_{50} - w_{50} \right)^{2}/w_{0}^2,
\label{eqn:residual_sum_of_squares}
\end{equation}
where $A$ and $w_{50}$ are the area and width 50\% of distance along the long axis of the mould calculated from results of RAPTOR or CONDOR for the selected mould shape, and $\bar{A}$ and $\bar{w}_{50}$ are the corresponding area and width measured from the experimental results. $A_0$ and $w_{0}$ are the uncontracted area and width respectively. The fit could be set up to optimise all the variable parameters in RAPTOR (which are $\Delta$, $\kappa_{\rm 3N}/\kappa_{\rm NN}$, and $\kappa_{\rm 3N}/\kappa_{\rm NN}$) however, in that case the fit is underdetermined. To avoid this problem, one or two parameters can be fixed to a specific value and only the remaining parameters optimised. This is justified since shear and bulk moduli are not very sensitive to the smallest $\kappa$ values in the contractile network, as shown in \ref{sec:appendixa}. We use a fixed ratio of $\kappa_{\rm 4N}/\kappa_{\rm 3N}=0.1$ for fitting purposes. For the case of corneal tissue, where only the width is known, we also fixed the ratio $\kappa_{\rm 3N}/\kappa_{\rm NN}=0.1$. For validation, the parameters determined from the fit can be input into the CONDOR simulation and compared against the RAPTOR output obtained using the same parameters.

The parameters obtained using this non-linear least squares fitting process are summarised in Table~\ref{tab:tissuesummary}, with a detailed description in Sec. \ref{sec:glial} (glial tissue), Sec. \ref{sec:fibroblasts} (fibroblast tissue), and Sec. \ref{sec:corneal} (corneal tissue). Table~\ref{tab:tissuesummary} also shows the relative width and area values obtained from both RAPTOR and CONDOR using the optimised parameters. As previously noted \cite{hague2019a}, CONDOR (and by extension RAPTOR) model parameters vary with cell type, as well as cell density and other factors such as passage (number of times that cells have been grown on and moved between flasks), and thus may even vary between experiments using the same types of hydrogels and cells \cite{hague2019a}. 

We also make use of an interpolated grid search to identify optimal parameters: We make CONDOR simulations for a grid of parameter values, read the contraction values from the simulation, and then interpolate between those values to make a smooth RSS curve in $\Delta$ using Eqn.~\ref{eqn:residual_sum_of_squares}. An example RSS curve arising from this procedure is shown in Fig~\ref{fig:contraction_curve_rss_x1171}. The minimum of the RSS curve is determined for each $\kappa_{\rm 3N}/\kappa_{\rm NN}$ on the grid, with the deepest minimum used to determine $\Delta$ and $\kappa_{\rm 3N}/\kappa_{\rm NN}$. We always set $\kappa_{\rm 4N}/\kappa_{\rm 3N}=0.1$. If only the width is known, we set $\kappa_{\rm 3N}/\kappa_{\rm NN}=0.1$. This method has the advantage of being separately applicable to both CONDOR and RAPTOR results, though it does requires CONDOR simulations to be pre-computed (which requires a lot more processing time than doing non-linear least squares using RAPTOR). Additionally, best fit values for $\kappa_{\rm 3N}/\kappa_{\rm NN}$ are limited to discrete values.

For both parameter optimisations, we have two experimental measures available. For this reason, we fix $\kappa_{\rm 4N}/\kappa_{\rm 3N} = 0.1$. We use value of $0.1$ as opposed to a midpoint since CONDOR simulations for that value have a range of shapes, including strong contraction for the $\Delta=0.9$ case (as demonstrated in Fig.~\ref{fig:varying_density_dogbone_simulation_d0x9} and Fig.~\ref{fig:varying_density_east_prediction_d0x9}). For the case of Corneal tissue for which we only know width contraction, so we also fix $\kappa_{\rm 3N}/\kappa_{\rm NN} = 0.1$.

\begin{table*}
    \centering
    \begin{tabular}{|c|c|c|c|c|c|c|c|c|c|c|c|}
    \hline
         Tissue & Mould Outline & \multicolumn{2}{c|}{Experimental} & \multicolumn{2}{c|}{Optimised} & Jaccard & \multicolumn{2}{c|}{RAPTOR} & \multicolumn{2}{c|}{CONDOR}\\
         & & \multicolumn{2}{c|}{} & \multicolumn{2}{c|}{Parameter} & Distance & \multicolumn{2}{c|}{} & \multicolumn{2}{c|}{} \\
         \hline
                  & & $\bar{A}/A_0$ & $\bar{w}_{50}/w_0$ & $\Delta$ & $\frac{\kappa_{\rm 3N}}{\kappa_{\rm NN}}$ & & $A/A_0$ & $w_{50}/w_0$ & $A/A_0$ & $w_{50}/w_0$ \\
         \hline
         & & & & & & & & & & \\
            Glial (G1) & \includegraphics[width=20mm]{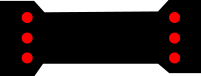} & 0.738 & 0.685 & 0.378 & 0.652 & 0.0540 & 0.687 & 0.633 & 0.699 & 0.683 \\[10pt]           
        Glial (G2) &  \includegraphics[height=12mm]{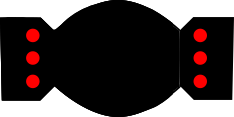} & 0.723 & 0.745 & 0.317 & 0.688 & 0.0818 & 0.713 & 0.719 & 0.722 & 0.702 \\[10pt]
        Glial (G3) &  \includegraphics[height=12mm]{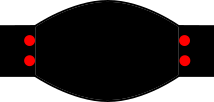} & 0.750 & 0.760 & 0.266 & 0.685 & 0.0610 & 0.743 & 0.761 & 0.761 & 0.771 \\[10pt]
        Fibroblast \cite{kostyuk2004} &  \includegraphics[height=12mm]{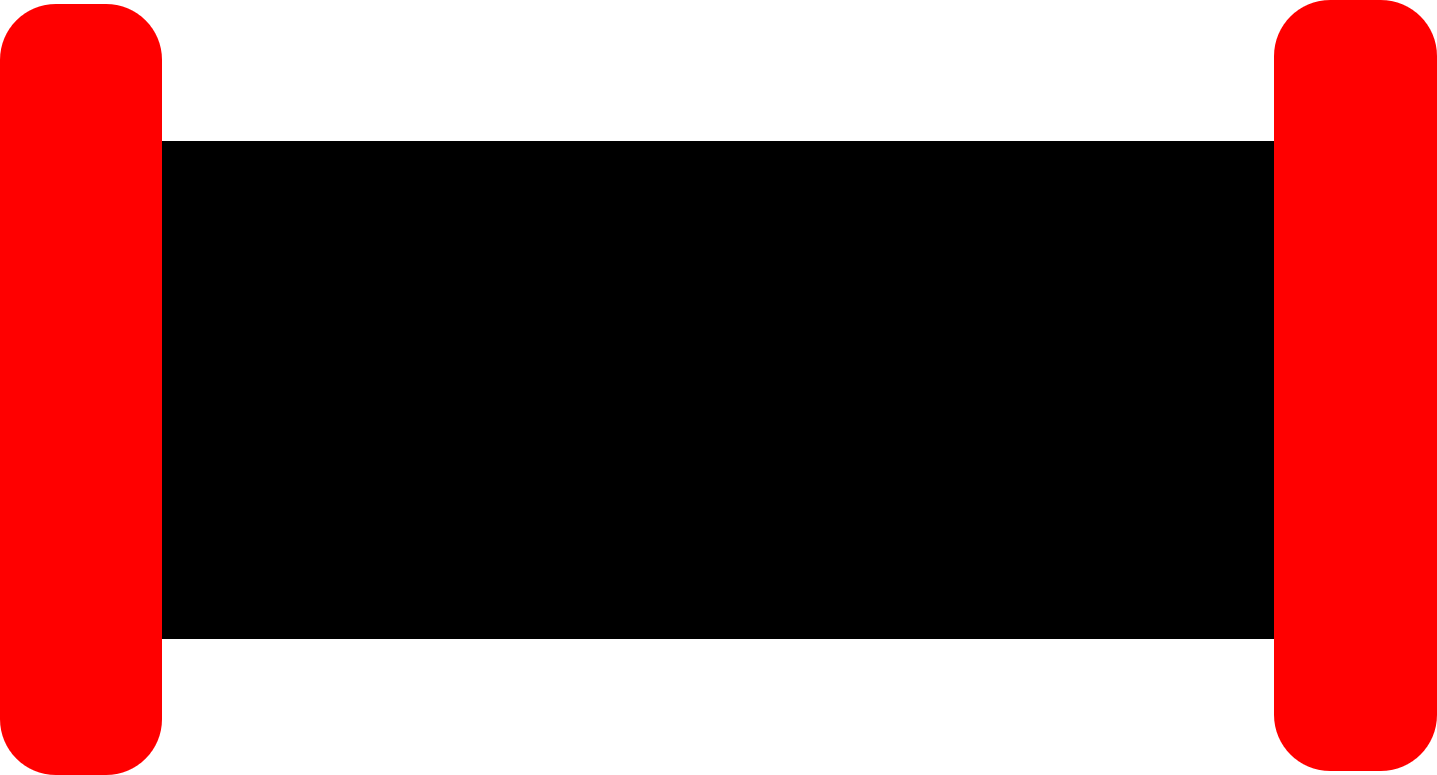} & 0.693 & 0.472 & 0.498 & 0.5353 &  0.0976 & 0.629 & 0.570 & 0.627 & 0.582 \\[10pt]
        Corneal \cite{mukhey2018} &  \includegraphics[height=12mm]{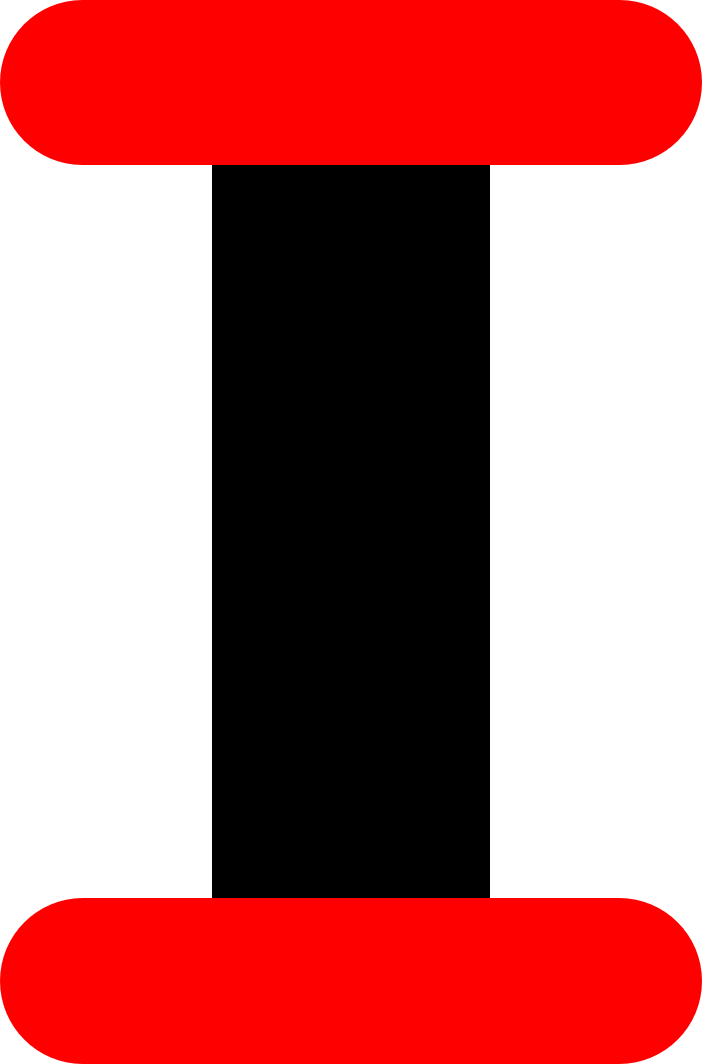} & - & 0.35 & 0.846 & 0.1 &  - & 0.512 & 0.492 & 0.421 & 0.311 \\[5pt]
        \hline
    \end{tabular}
    \caption{A summary of widths and areas from experiments \cite{kostyuk2004,mukhey2018,orourke2015}, and model parameters determined from RAPTOR using non-linear least squares (marked prediction) and comparative widths and areas obtained from operating CONDOR and RAPTOR with the optimised parameters. $\kappa_{\rm 4N}/\kappa_{\rm 3N}=0.1$ throughout. Unavailable values are labelled with a hyphen. The moulds are shown in outline, with tethers in red and the mould depression in black.
    }
    \label{tab:tissuesummary}
\end{table*}

\begin{table*}
    \centering
    \begin{tabular}{|c|c|c|c|c|c|c|c|c|c|c|}
    \hline
         Tissue & Mould Outline & \multicolumn{2}{c|}{Experimental} & \multicolumn{2}{c|}{Optimised} & Jaccard & \multicolumn{2}{c|}{CONDOR} & \multicolumn{2}{c|}{RAPTOR}\\
         & & \multicolumn{2}{c|}{} & \multicolumn{2}{c|}{Parameter} & Distance & \multicolumn{2}{c|}{} & \multicolumn{2}{c|}{} \\
         \hline
         & & $\bar{A}/A_0$ & $\bar{w}_{50}/w_0$ & $\Delta$ & $\frac{\kappa_{\rm 3N}}{\kappa_{\rm NN}}$ & & $A/A_0$ & $w_{50}/w_0$ & $A/A_0$ & $w_{50}/w_0$ \\
         \hline
         & & & & & & & & & & \\
        Glial (G1) & \includegraphics[width=20mm]{1T_X1171-A.png} & 0.814 & 0.797 & 0.38 & 0.1 & 0.0741 & 0.664 & 0.683 & 0.671 & 0.633 \\[10pt]
        Glial (G2) &  \includegraphics[height=12mm]{2T_X1177-A.png} & 0.863 & 0.874 & 0.36 & 0.3 & 0.0911 & 0.694 & 0.675 & 0.677 & 0.684 \\[10pt]
        Glial (G3) &  \includegraphics[height=12mm]{3T_X1178-A.png} & 0.750 & 0.759 & 0.28 & 0.3 & 0.0759 & 0.780 & 0.789 & 0.722 & 0.725\\[10pt]
        Fibroblast \cite{kostyuk2004} &  \includegraphics[height=12mm]{4T_kostyuk.png} & 0.693 & 0.472 & 0.66 & 0.7 & 0.0780 & 0.539 & 0.494 & 0.546 & 0.481 \\[10pt]
        Corneal \cite{mukhey2018} &  \includegraphics[height=12mm]{5T_imould_east.png} & - & 0.35 & 0.90 & 0.1 & - & 0.398 & 0.295 & 0.496 & 0.492 \\[5pt]

        \hline
    \end{tabular}
    \caption{As Table \ref{tab:tissuesummary}, but with model parameters determined from CONDOR using an RSS grid search method, and comparative widths and areas obtained from operating CONDOR and RAPTOR with the same optimised parameters. $\kappa_{\rm 4N}/\kappa_{\rm 3N}=0.1$ throughout.
    }
    \label{tab:tissuesummary_2}
\end{table*}

\subsubsection{Glial cultures}
\label{sec:glial}

In this section, comparisons are made with cultured glial tissue (glial cell populated hydrogels in tethered moulds). Glial cells play an important role in the nervous system, and direct the growth and support of neurons. Nerve repair guides are an application of highly aligned glial tissue \cite{georgiou2013,phillips2021}. We have validated against lab-grown glial tissue for several tethered moulds. The work in this section also augments previous validation of CONDOR against glial tissues grown in tethered moulds \cite{hague2019a}.

\begin{figure*}[h]
\centering
{
\includegraphics[width=0.8\textwidth]{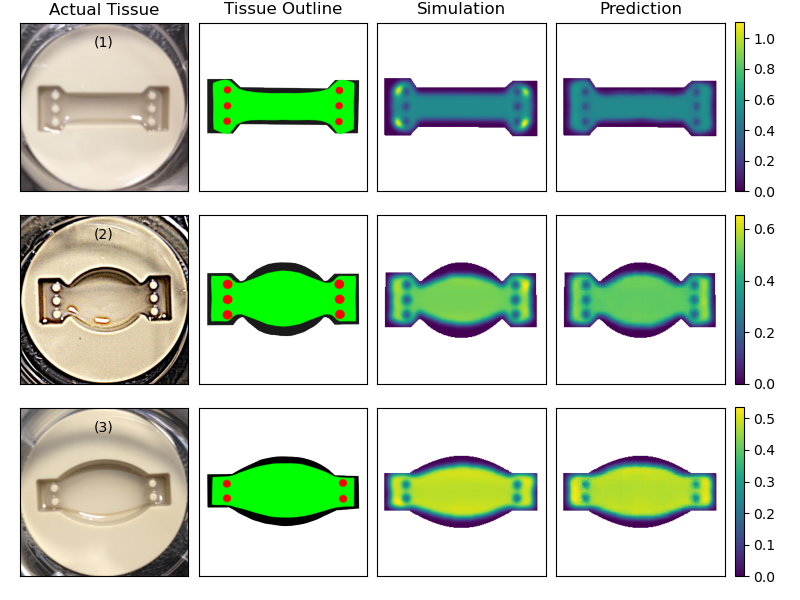}
\caption{Column 1: Comparison between experimental glial tissue samples grown in three different moulds (glial G1, glial G2 and glial G3). Column 2: the traced outline. Columns 3 and 4: the best fit CONDOR simulation and RAPTOR predictions. Good agreement is found between the simulations, prediction and tissue.
}

\label{fig:real_comparison_x1171}
\label{fig:real_comparison_x1177}
\label{fig:real_comparison_x1178}
}
\end{figure*}

\begin{figure*}[h]
\centering
{
\begin{subfigure}[t]{0.8\textwidth}
\caption{Simulation}
\includegraphics[width=0.8\textwidth]{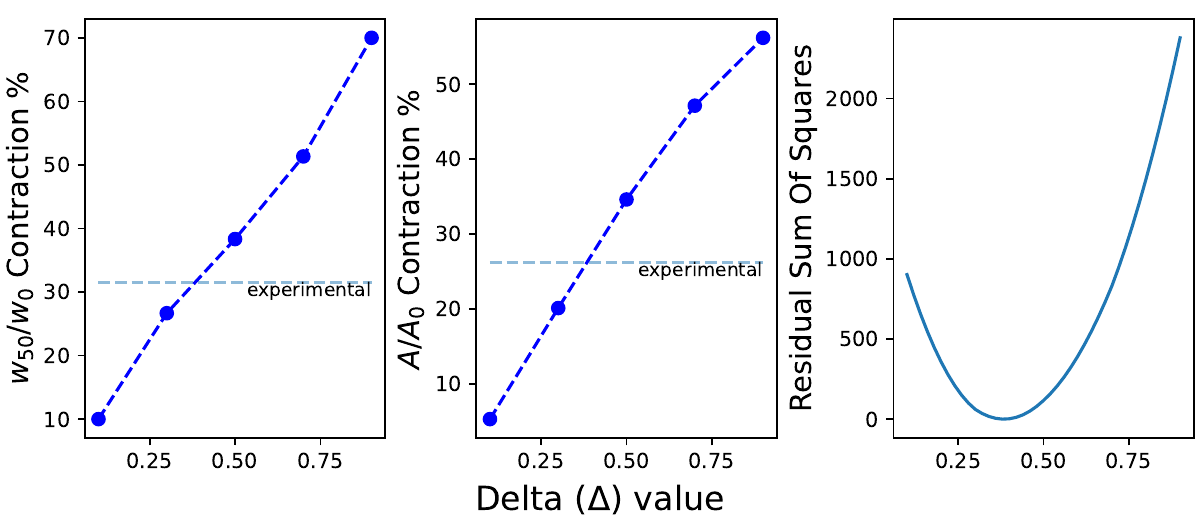}    
\end{subfigure}
\begin{subfigure}[t]{0.8\textwidth}
\caption{Prediction}
\includegraphics[width=0.8\textwidth]{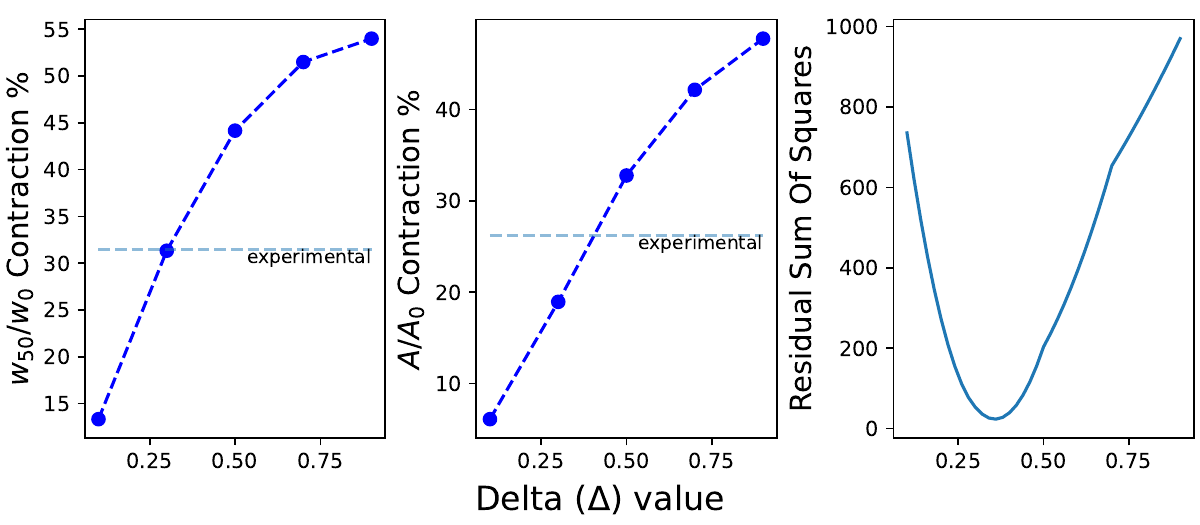}
\end{subfigure}
\caption{Central width and area contraction curves for glial tissue (mould G1) for parameters $\kappa_{\rm 3N}/\kappa_{\rm NN}=0.5$ and $\kappa_{\rm 4N}/\kappa_{\rm 3N}=0.1$, alongside the residual sum of squares.
}

\label{fig:contraction_curve_rss_x1171}

}
\end{figure*}

Comparisons of laboratory grown glial tissue \cite{orourke2015}, CONDOR simulation and RAPTOR prediction are shown for 3D glial cultures in three different moulds in Fig. \ref{fig:real_comparison_x1171}. Qualitative agreement between the shapes is very good in all cases. Good agreement is also found for the width and area ratios determined from RAPTOR and CONDOR and those of the experimentally cultured tissue (see Tables \ref{tab:tissuesummary} and \ref{tab:tissuesummary_2}).

All cases shown in Fig. \ref{fig:real_comparison_x1171} show glial tissue grown under similar conditions. Thus, it would be expected that they possess similar contractile properties and CONDOR parameters. Moderate variation is found in the parameters determined from both optimisation processes for the three moulds (Tables~\ref{tab:tissuesummary} and \ref{tab:tissuesummary_2}). The average $\Delta = 0.320 \pm 0.032$ for RAPTOR fits and $0.340 \pm 0.031$ for CONDOR. Optimised values of $\kappa_{\rm 3N}/\kappa_{\rm NN}$ are of similar magnitude. Jaccard distances relating predictions and simulations to experimental tracings indicate a good match between simulation, prediction and experiment, with all distances less than $0.1$ (see Tables~\ref{tab:tissuesummary} and \ref{tab:tissuesummary_2} for values).

We note that if $\kappa_{\rm 4N}/\kappa_{\rm 3N}$ is not constrained, the resulting parameter values for X1171 are $[\Delta, \kappa_{\rm 3N}/\kappa_{\rm NN},\kappa_{\rm 4N}/\kappa_{\rm 3N}]$ $ =  [0.321,0.611,0.545]$, 
for X1177 parameters are $[0.312,0.611,0.543]$ 
 and for X1178 $[0.272,0.636, 0.573]$. These have a lower variance than the constrained cases, but the fitting problem is technically underdetermined.

Overall, however, both fit processes were able to determine model parameters for CONDOR and RAPTOR methods for further simulation and design of tethered moulds. All optimised parameters led to results that match experimental width and area measurements. Good agreement is found between experiment, biophysical simulation and machine learning predictions for the shape of the 3D glial cultures considered here.

\subsubsection{Fibroblast cultures}
\label{sec:fibroblasts}

The results from RAPTOR and CONDOR for fibroblasts were compared to relative tissue widths and areas derived from images in Ref. \cite{kostyuk2004}. Fibroblasts synthesise collagen and play in important role in tissue healing and repair. Fibroblast tissue is relatively contractile, an important feature related to its role in wound closure.

Using image data from Ref. \cite{kostyuk2004}, we estimated width and area contractions (recorded in Tables~\ref{tab:tissuesummary} and \ref{tab:tissuesummary_2}) that were subsequently input into our parameter optimisation. The resulting parameter values from the least-squares fit method (Table~\ref{tab:tissuesummary}) are $\Delta = 0.49814$ and $\kappa_{\rm 3N}/\kappa_{\rm NN}=0.5353$, while for the RSS grid search method (Table~\ref{tab:tissuesummary_2}) are $\Delta = 0.66$ and $\kappa_{\rm 3N}/\kappa_{\rm NN}=0.7$. Both $\Delta$ values are much higher than those for glial tissues indicating larger contractile forces for fibroblast cultures.

\begin{figure*}[h]
\centering
{
\includegraphics[width=0.9\textwidth]{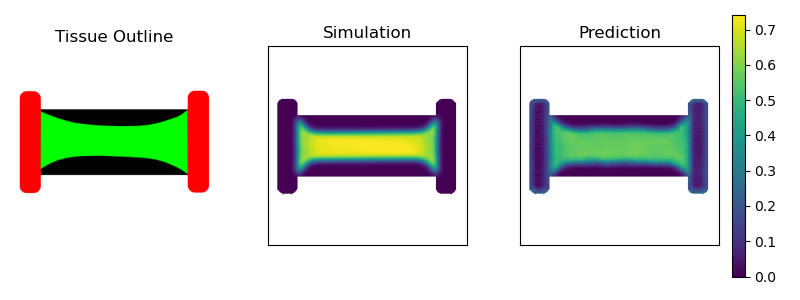}
\caption{Best matches between the RAPTOR prediction and the cultured fibroblast tissue sample discussed in Ref. \cite{kostyuk2004}. Widths, areas and $\Delta$ values can be found in Tables \ref{tab:tissuesummary} and \ref{tab:tissuesummary_2}. The tissue outline was established by using InkScape \cite{inkscape} to trace around the outline of the tissue in Figure 1b of Ref. \cite{kostyuk2004}.}
\label{fig:real_comparison_kostyuk}
}
\end{figure*}

Figure~\ref{fig:real_comparison_kostyuk} shows the comparison between the CONDOR simulation and the RAPTOR prediction for for the optimised parameters. Qualitative agreement between the shapes is very good, though we note a small difference between CONDOR and RAPTOR regarding area and width ratios. Again, the both fitting processes have found suitable CONDOR / RAPTOR model parameters for further biophysical design and simulation.

Quantitative comparison between the tracing and tissue shape from simulation and prediction (calculated from $N(P>0.2)$) also shows good agreement. The Jaccard distance between the tracing and the RAPTOR prediction is $0.0669$. The Jaccard distance between the CONDOR simulation and traced outline of the experiment is $0.0781$.

\subsubsection{Corneal cultures}
\label{sec:corneal}

We use the width provided in Ref. \cite{mukhey2018} to compare RAPTOR predictions with experimental results for engineered tissue containing corneal stromal cells. Corneal tissue is an important part of the eye, and high alignment of cells and particularly matrix is important for the optical properties of the cornea.  Ref. \cite{mukhey2018} does not provide a high resolution image of the final contracted tissue from which we can make a direct comparison of the shapes, but does provide measurements of width contraction. We note that the same mould was used for glial tissue growth in Ref. \cite{east2010}.

We use the width value from Ref. \cite{mukhey2018} in our parameter fit to obtain values of $\Delta$, while fixing $\kappa_{\rm 3N}/\kappa_{\rm NN}$ and $\kappa_{\rm 4N}/\kappa_{\rm 3N}$ to $0.1$ since only width was reported in Ref. \cite{mukhey2018}. The least-squares fitting method returns $\Delta=0.8457$, while the RSS grid search returns $\Delta=0.90$.

\begin{figure*}[h]
\centering
{
\includegraphics[width=0.9\textwidth]{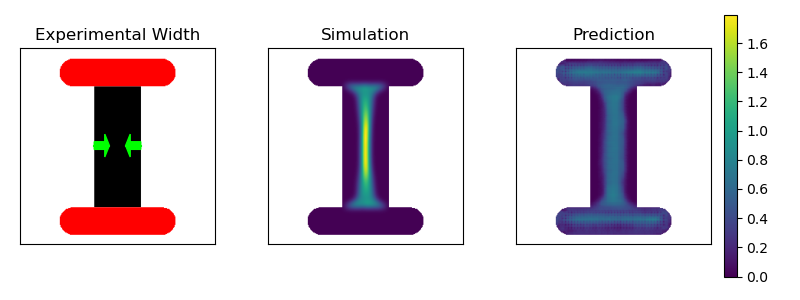}
\caption{Best matches between the RAPTOR prediction and the cultured corneal tissue sample discussed in Ref. \cite{mukhey2018}. Widths, areas and $\Delta$ values can be found in Tables \ref{tab:tissuesummary} and  \ref{tab:tissuesummary_2}. Since only the width is known, the fits were made with both fixed $\kappa_{\rm 3N}/\kappa_{\rm NN} = 0.1$ and $\kappa_{\rm 4N}/\kappa_{\rm 3N}=0.1$. Note that no images of the fully contracted tissue were provided in Ref. \cite{mukhey2018}. The two arrows in the plot indicate the extent of the width of the corneal tissue sample as reported in Ref. \cite{mukhey2018}.}
\label{fig:real_comparison_mukhey}
}
\end{figure*}

Figure~\ref{fig:real_comparison_mukhey} shows the comparison between the results of the CONDOR simulation and RAPTOR prediction when using the least-squares fit results. There is a small difference between overall density and the central width predicted by each method, with the machine learning model predicting slightly less contraction across the centre of the tissue. This likely arises from the challenges of making RAPTOR predictions for high $\Delta$ and small $\kappa$ values discussed in Section \ref{sec:raptorresults}.

\subsection{Limitations}

\subsubsection{Highly contractile tissue}

The biggest differences between the RAPTOR predictions and CONDOR output are for cases with high contraction. For example, comparison between width and area contractions in Figure~\ref{fig:contraction_curves_dogbone} show that the contraction predicted from RAPTOR differs primarily with the prediction for CONDOR in cases with low $\kappa_{\rm 3N}/\kappa_{\rm NN}$, low $\kappa_{\rm 4N}/\kappa_{\rm 3N}$ and high $\Delta$. Comparisons between CONDOR and RAPTOR in Figure \ref{fig:contraction_curves_east} shows similar limitations: width and area contraction is underestimated by RAPTOR, with bigger differences for large $\Delta$ and small $\kappa$. We believe that this difference occurs because tissue shapes are not strongly dependent of $\kappa_{\rm 3N}/\kappa_{\rm NN}$ and $\kappa_{\rm 4N}/\kappa_{\rm NN}$ in most cases. This means that there are limited examples of highly contracted cases in the training data set, which can be found only for the smallest $\kappa$ and largest $\Delta$ values.

This is a challenge for making machine learning predictions for tissue cultures of tenocytes and myoblasts. Representative examples of artificial tendon and muscle tissues grown in tethered moulds can be found in Refs. \cite{garvin2003a} and \cite{capel2019a}. Both types of tissue are highly contractile. The maximum $\Delta = 0.95$ is returned from the parameter search in both cases. However, spring constant values ($\kappa_{\rm 3N}$ and $\kappa_{\rm 4N}$) returned from the search are within the region of parameter space where RAPTOR predictions do not match CONDOR well (as discussed earlier).  Discussion of the use of CONDOR to model these two cell types is reserved for a subsequent publication.

\subsubsection{Time dependence}

We also note that the model does not directly include time dependence. Several different timescales can be found in cultivated tissues. Cells divide and move within the ECM over a period of many day. ECM remodelling occurs on timescales between hours and days. Finally, the reaction of the matrix to ECM remodelling is essentially instantaneous in this context.

One could calculate the dynamics of the model by deriving equations of motion from Eqn. \ref{eqn:hamiltonian}. However, this would simulate the immediate reaction of the matrix to ECM remodelling by cells, which is essentially instantaneous and not relevant to biological time scales. 

Rudimentary time dependence could be established by using time dependent parameters $\Delta(t)$ and $\kappa(t)$. These could be locally dependent on the cell orientation. To provide full insight into time dependent processes, a set of rules would need to be established to simulate the ECM remodelling, which would require a significant extension to the model and is reserved for future work. 

\section{Summary and conclusions}
\label{sec:summary}

In summary, we have presented a machine learning approach for rapid prediction of tissue organisation based on the \texttt{pix2pix} model for a variety of different cell types, and validated the approach for various cultured tissues. The key difference between this and previous work with CONDOR-ML \cite{andrews2023} is that CONDOR model parameters can be varied in the new RAPTOR approach, allowing predictions for a variety of tissue types. Different cell types and their densities within 3D cultured tissues are associated with different model parameters, so this feature of RAPTOR is essential to make predictions for a range of cell types, cell densities, ECM types or the effects of other factors such as enzymes applied to cultured tissue. We found good agreement between RAPTOR and CONDOR predictions and contractions of glial, fibroblast and corneal tissue grown in the laboratory.

RAPTOR and CONDOR predictions agree across a large part of the model parameter space. However, the parameter space grids calculated for both CONDOR and RAPTOR also show that for cases with both low $\kappa$ and high $\Delta$, RAPTOR was not able to precisely represent the highly contracted cultured tissues predicted by CONDOR. We believe that lower frequency of cases with strong contraction in the training data set cause the lower contractions predicted by RAPTOR. A machine learning model might be trained using a data set specialised to this small region of parameter space to make more accurate predictions for those cases.

We established that RAPTOR can be used to make rapid determination of the CONDOR and RAPTOR model parameters to represent particular cultured tissue types. Two approaches to minimise residual sum of squares using RAPTOR and CONDOR were developed to establish model parameters by matching the areas and widths of RAPTOR/CONDOR results to tissues cultured in the laboratory using tethered moulds. The method using RAPTOR as part of a least squares fitting routine is much faster than grid searches using CONDOR only. The parameters (from either method) could then be suitable to guide computational design of moulds and tethers using CONDOR or RAPTOR.

The speed of the RAPTOR method makes a near real-time design process possible. It also makes automated design through computational intelligence techniques such as evolutionary strategies a possibility. Exploration of this potential will form part of a future paper \cite{andrews2025b}.

\section*{Acknowledgments}

We acknowledge support from the STFC Impact Accelerator Account 2022-2023.

\bibliographystyle{plain}
\bibliography{references}

\appendix

\section{Moduli for face-center-cubic contractile network}
\label{sec:appendixa}

To calculate bulk and shear moduli for a contractile network, small changes to the network need to me made. In the following, we assume bonds radiating out from the origin to simplify the calculations, however the resulting expressions are general. For a single bond anchored at the origin, the difference in energy due to a small change in position, $\etavec$ at the other end of the bond is:
\begin{eqnarray}
    \delta E & = & \frac{\kappa}{2}(|\lvec+\etavec|-|\lvec|)^2\\
    & = & \frac{\kappa}{2}(\sqrt{(\lvec+\etavec)\cdot(\lvec+\etavec)}-\sqrt{\lvec\cdot\lvec})^2\nonumber\\
    & \approx & \frac{\kappa}{2}(\sqrt{\lvec\cdot\lvec+2\etavec\cdot\lvec+\mathcal{O}(\eta^2)}-\sqrt{\lvec\cdot\lvec})^2\nonumber\\
        & = & \frac{\kappa}{2}(\sqrt{\lvec\cdot\lvec}\sqrt{1+2\etavec\cdot\lvec/\lvec\cdot\lvec+\mathcal{O}(\delta^2)}-\sqrt{\lvec\cdot\lvec})^2\nonumber\\
    & \approx & \frac{\kappa}{2}\left(\sqrt{\lvec\cdot\lvec} + \frac{1}{2}2\etavec\cdot\lvec/\sqrt{\lvec\cdot\lvec} - \sqrt{\lvec\cdot\lvec}\right)^2\nonumber\\
    & = & \frac{\kappa}{2}\left(\etavec\cdot\hat{\lvec}\right)^2
\end{eqnarray}

In three dimensions with cubic symmetry, the free energy density of a continuum elasticity model can be related to the strain via \cite{boal},
\begin{eqnarray}
    \Delta\mathcal{F} = & & (K_V/2)(u_{xx}+u_{yy}+u_{zz})^2\nonumber\\
    & & +(\mu'/3)\left[(u_{xx}-u_{yy})^2+(u_{xx}-u_{zz})^2+(u_{yy}-u_{zz})^2\right]\nonumber\\
    & &+2\mu_{s}(u_{xy}^2+u_{xz}^2+u_{yz}^2)
\end{eqnarray}
For simple isotropic compression and no shear, the network energy can be related to the term in the continuum free energy with coefficient $K_{V}$. For simple shear of $u_{xy}$ only, $\mu_{s}$ can be determined. Finally, for $u_{xx}=u_{yy}=-u_{zz}/2$ and no shear, the pure shear modulus can be determined.  A set of suitable deformations can be found in Table \ref{tab:freeenchange}. Note, only half of all bonds are in the table to avoid double counting. Those deformations need to be compared with the free energy for a continuum material (see e.g. Ref. \cite{boal}).

The effects of different matrix properties are essentially shown in Figs. \ref{fig:varying_density_dogbone_simulation_d0x5}, \ref{fig:varying_density_dogbone_simulation_d0x9}, \ref{fig:varying_density_east_prediction_d0x5} and \ref{fig:varying_density_east_prediction_d0x9}, since the relative size of shear and bulk moduli changes with the different $\kappa$. For the contractile networks here, shear modulus is dominated by $\kappa_{\rm NN}$ when $\kappa_{\rm 4N}/\kappa_{\rm NN}<~ 0.1$ and bulk modules for $\kappa_{\rm 4N}/\kappa_{\rm NN}$ and $\kappa_{\rm 4N}/\kappa_{\rm 3N}<~ 0.1$, $\kappa_{\rm 3N}/\kappa_{\rm NN}<~ 1$, which leads to the regions in those Figures where the tissue properties do not strongly change.

\begin{table*}
    \begin{tabular}{c|cc|cc|cc}
    $\lvec/a$ & $\etavec$ (sim. shear) & $\delta E_{s}$ & $\etavec$ (pure shear) & $\delta E_{p}$ & $\etavec$ (compression) & $\delta E_{c}$ \\
    \hline
\multicolumn{7}{c}{NN, bond-length $a/\sqrt{2}$} \\
    \hline
    $\frac{1}{2}(\exvec+\eyvec)$ & 0 & 0 & $\delta(\exvec + \eyvec)/2$& $ \frac{\kappa \delta^2}{4}$ & $-\delta (\exvec+\eyvec)/2$ & $ \frac{\kappa \delta^2}{4}$ \\
    $\frac{1}{2}(-\exvec+\eyvec)$ & $0$ & $0$ & $-\delta (\exvec - \eyvec)/2 $ & $ \frac{\kappa \delta^2}{4}$ & $\delta (\exvec - \eyvec)/2 $ & $ \frac{\kappa \delta^2}{4}$ \\
        $\frac{1}{2}(\exvec+\ezvec)$ & $\delta \exvec/2$ & $\frac{\kappa\delta^2}{16}$ & $\delta (\exvec/2 - \ezvec) $  & $\frac{\kappa \delta^2}{16}$ & $-\delta (\exvec + \ezvec)/2$ & $ \frac{\kappa \delta^2}{4}$ \\
    $\frac{1}{2}(-\exvec+\ezvec)$ & $\delta \exvec/2$  & $\frac{\kappa\delta^2}{16}$ & $-\delta (\exvec/2 + \ezvec) $ & $\frac{\kappa \delta^2}{16}$ & $\delta (\exvec - \ezvec)/2 $ & $ \frac{\kappa \delta^2}{4}$ \\
        $\frac{1}{2}(\eyvec+\ezvec)$ & $\delta \exvec/2$  & 0 & $\delta (\eyvec/2 - \ezvec) $  & $\frac{\kappa \delta^2}{16}$ & $-\delta (\ezvec + \eyvec)/2$ & $ \frac{\kappa \delta^2}{4}$ \\
    $\frac{1}{2}(-\eyvec+\ezvec)$ & $\delta \exvec/2$  & $0$ & $-\delta (\eyvec/2 + \ezvec) $ & $\frac{\kappa \delta^2}{16}$ & $\delta (\ezvec -  \eyvec)/2 $ & $ \frac{\kappa \delta^2}{4}$ \\
    
    \hline
    \multicolumn{7}{c}{3N, bond-length $a$} \\
    \hline
    $\exvec$ & 0 & 0 & $\delta \exvec$& $\frac{\kappa \delta^2}{2}$ & $-\delta \exvec$ & $\frac{\kappa \delta^2}{2}$  \\
    $\eyvec$ & $0$ & 0 & $\delta \eyvec$ & $\frac{\kappa \delta^2}{2}$ & $-\delta \eyvec$ & $\frac{\kappa \delta^2}{2}$\\
    $\ezvec$ & $\delta \exvec$ & 0 & $-2\delta \ezvec$ & $\kappa 2\delta^2$ & $-\delta \ezvec$ & $\frac{\kappa \delta^2}{2}$\\

    \hline
    \multicolumn{7}{c}{4N, bond-length $\sqrt{3}a/\sqrt{2}$} \\
    \hline
    $\exvec+\frac{1}{2}[\eyvec+\ezvec]$ & $\delta \exvec/2$ & $\frac{\kappa_{\rm 4N}\delta^2}{12}$ & $\delta \exvec + \delta \eyvec/2-\delta\ezvec $  & $\frac{\kappa 3\delta^2}{16}$ & $-\delta (\exvec + \eyvec/2+\ezvec/2)$ & $\frac{3 \kappa_{\rm 4N} \delta^2}{4}$ \\

        $\exvec-\frac{1}{2}[\eyvec+\ezvec]$ & $\delta \exvec/2$ & $\frac{\kappa_{\rm 4N}\delta^2}{12}$ & $\delta \exvec - \delta \eyvec/2+\delta\ezvec $  & $\frac{\kappa 3\delta^2}{16}$ & $-\delta (\exvec - \eyvec/2-\ezvec/2)$ & $\frac{3 \kappa_{\rm 4N} \delta^2}{4}$ \\

            $\exvec+\frac{1}{2}[\eyvec-\ezvec]$ & $\delta \exvec/2$ & $\frac{\kappa_{\rm 4N}\delta^2}{12}$ & $\delta \exvec + \delta \eyvec/2+\delta\ezvec $  & $\frac{\kappa 3\delta^2}{16}$ & $-\delta (\exvec + \eyvec/2-\ezvec/2)$ & $\frac{3 \kappa_{\rm 4N} \delta^2}{4}$ \\

               $\exvec-\frac{1}{2}[\eyvec-\ezvec]$ & $\delta \exvec/2$ & $\frac{\kappa_{\rm 4N}\delta^2}{12}$ & $\delta \exvec - \delta \eyvec/2-\delta\ezvec $  & $\frac{\kappa 3\delta^2}{16}$ & $-\delta (\exvec - \eyvec/2+\ezvec/2)$ & $\frac{3 \kappa_{\rm 4N} \delta^2}{4}$ \\
                
                $\eyvec+\frac{1}{2}[\exvec+\ezvec]$ & $\delta \exvec/2$ & $\frac{\kappa_{\rm 4N}\delta^2}{48}$ & $\delta \exvec/2 + \delta \eyvec-\delta\ezvec $  & $\frac{\kappa 3\delta^2}{16}$ & $-\delta (\eyvec -\exvec/2+\ezvec/2)$ & $\frac{3 \kappa_{\rm 4N} \delta^2}{4}$ \\

                $\eyvec-\frac{1}{2}[\exvec+\ezvec]$ & $\delta \exvec/2$ & $\frac{\kappa_{\rm 4N}\delta^2}{48}$ & $-\delta \exvec/2 + \delta \eyvec+\delta\ezvec $  & $\frac{\kappa 3\delta^2}{16}$ & $-\delta (\eyvec - \exvec/2-\ezvec/2)$ & $\frac{3 \kappa_{\rm 4N} \delta^2}{4}$ \\

                $\eyvec+\frac{1}{2}[\exvec-\ezvec]$ & $\delta \exvec/2$ & $\frac{\kappa_{\rm 4N}\delta^2}{48}$ & $\delta \exvec/2 + \delta \eyvec+\delta\ezvec $  & $\frac{\kappa 3\delta^2}{16}$ & $-\delta (\eyvec + \exvec/2-\ezvec/2)$ & $\frac{3 \kappa_{\rm 4N} \delta^2}{4}$ \\

                $\eyvec-\frac{1}{2}[\exvec-\ezvec]$ & $\delta \exvec/2$ & $\frac{\kappa_{\rm 4N}\delta^2}{48}$ & $-\delta \exvec/2 + \delta \eyvec-\delta\ezvec $  & $\frac{\kappa 3\delta^2}{16}$ & $-\delta (\eyvec - \exvec/2+\ezvec/2)$ & $\frac{3 \kappa_{\rm 4N} \delta^2}{4}$ \\
                
                $\ezvec+\frac{1}{2}[\exvec+\eyvec]$ & $\delta \exvec$ & $\frac{\kappa_{\rm 4N}\delta^2}{12}$ & $\delta \exvec/2 + \delta \eyvec/2-2\delta\ezvec $  & $\frac{\kappa 3\delta^2}{4}$ & $-\delta (\ezvec + \exvec/2+\eyvec/2)$ & $\frac{3 \kappa_{\rm 4N} \delta^2}{4}$ \\

                $\ezvec-\frac{1}{2}[\exvec+\eyvec]$ & $\delta \exvec$ & $\frac{\kappa_{\rm 4N}\delta^2}{12}$ & $-\delta \exvec/2 - \delta \eyvec/2-2\delta\ezvec $  & $\frac{\kappa 3\delta^2}{4}$ & $-\delta (\ezvec - \exvec/2-\eyvec/2)$ & $\frac{3 \kappa_{\rm 4N} \delta^2}{4}$ \\
                
                $\ezvec+\frac{1}{2}[\exvec-\eyvec]$ & $\delta \exvec$ & $\frac{\kappa_{\rm 4N}\delta^2}{12}$ & $\delta \exvec/2 - \delta \eyvec/2-2\delta\ezvec $  & $\frac{\kappa 3\delta^2}{4}$ & $-\delta (\ezvec + \exvec/2-\eyvec/2)$ & $\frac{3 \kappa_{\rm 4N} \delta^2}{4}$ \\

                $\ezvec-\frac{1}{2}[\exvec-\eyvec]$ & $\delta \exvec$ & $\frac{\kappa_{\rm 4N}\delta^2}{12}$ & $-\delta \exvec/2 - \delta \eyvec/2 -2\delta\ezvec $  & $\frac{\kappa 3\delta^2}{4}$ & $-\delta (\ezvec - \exvec/2+\eyvec/2)$ & $\frac{3 \kappa_{\rm 4N} \delta^2}{4}$ \\
    
    \hline
       \multicolumn{7}{c}{Next-next-next-next-nearest neighbours (5N), bond-length $a\sqrt{2}$}\\ 
    \hline
     $\exvec+\eyvec$ & 0 & 0 & $\delta \exvec + \delta \eyvec $  & $\kappa_{\rm 4N} \delta^2$ & $-\delta (\exvec + \eyvec)$ & $\kappa_{\rm 4N} \delta^2$ \\
    $-\exvec+\eyvec$ & $0$ & $0$ & $-\delta \exvec + \delta \eyvec $ & $\kappa_{\rm 4N} \delta^2$ & $\delta \exvec - \delta \eyvec $ & $\kappa_{\rm 4N} \delta^2$ \\
        $\exvec+\ezvec$ & $\delta \exvec$ & $\frac{\kappa_{\rm 5N}\delta^2}{4}$ & $\delta \exvec - 2\delta \ezvec $  & $\frac{\kappa_{\rm 5N}\delta^2}{4}$ & $-\delta (\exvec + \ezvec)$ & $ \kappa_{\rm 4N} \delta^2$ \\
    $-\exvec+\ezvec$ & $\delta \exvec$  & $\frac{\kappa_{\rm 5N}\delta^2}{4}$ & $-\delta \exvec - 2\delta \ezvec $ & $\frac{\kappa_{\rm 5N}\delta^2}{4}$ & $\delta \exvec - \delta \ezvec $ & $ \kappa_{\rm 4N} \delta^2$ \\
        $\eyvec+\ezvec$ & $\delta \exvec$  & 0 & $\delta \eyvec - 2\delta \ezvec $  & $\frac{\kappa_{\rm 5N}\delta^2}{4}$ & $-\delta (\ezvec + \eyvec)$ & $ \kappa_{\rm 4N} \delta^2$ \\
    $-\eyvec+\ezvec$ & $\delta \exvec$  & $0$ & $-\delta \eyvec - 2\delta \ezvec $ & $\frac{\kappa_{\rm 5N}\delta^2}{4}$ & $\delta \ezvec - \delta \eyvec $ & $\kappa_{\rm 4N} \delta^2$ \\

         \hline
     \multicolumn{7}{c}{Summary of moduli}\\
     \hline
    $\delta E_{\rm tot}/V_{\rm v}$ & \multicolumn{2}{c|}{$\frac{\delta^2}{a^3}\left(\frac{\kappa_{\rm NN}}{2}+3\kappa_{\rm 4N}+2\kappa_{\rm 5N}\right)$} & \multicolumn{2}{c|}{$\frac{\delta^2}{a^3}\left(3\kappa_{\rm NN}+12\kappa_{\rm 3N}+6\kappa_{\rm 4N}+12\kappa_{\rm 5N}\right)$} & \multicolumn{2}{|c}{$\frac{\delta^2}{a^3}\left(6\kappa + 6\kappa_{\rm 3N}+36\kappa_{\rm 4N}+24\kappa_{\rm 5N}\right)$} \\
     \hspace{5mm}$(V_{\rm v}=a^{3}/4)$ &  & & & & & \\
   & & & & & &  \\
    Strain & \multicolumn{2}{c|}{$u_{xz} = \delta/a$} & \multicolumn{2}{c|}{$u_{xx} = u_{yy}=\delta/a$, $u_{zz}=-2\delta/2$} & \multicolumn{2}{c}{$u_{xx}=u_{yy}=u_{zz}=-\delta/a$} \\ 
        Free energy dens. & \multicolumn{2}{c|}{$2\mu_{s}u_{xz}^2$} & \multicolumn{2}{c|}{$ \frac{\mu_{p}}{3}\left[(u_{xx}-u_{zz})^2+(u_{yy}-u_{zz})^2\right]$} & \multicolumn{2}{c}{$\frac{K_{A}}{2}(u_{xx}+ u_{yy}+u_{zz})^2$} \\
     & \multicolumn{2}{c|}{$\equiv 2\mu_{s}\delta^2/a^2$} & \multicolumn{2}{c|}{$\equiv 6\mu_{p}\delta^2/a^2$}  &  \multicolumn{2}{c}{$\equiv 9K_{a}\delta^2/2a^2$}\\
    & & & & & &  \\
   Modulus & \multicolumn{2}{|c}{$\mu_{s} =\frac{1}{a} (\frac{\kappa_{\rm NN}}{16}+\frac{3\kappa_{\rm 4N}}{8}+\frac{\kappa_{\rm 5N}}{4})$} & \multicolumn{2}{|c}{$\mu_{p} = \frac{1}{a}\left(\frac{\kappa_{\rm NN}}{8}+\frac{\kappa_{\rm 3N}}{2}+\frac{\kappa_{\rm 4N}}{4}+\frac{\kappa_{\rm 5N}}{2}\right)$} & \multicolumn{2}{|c}{$K_{V}=\frac{1}{a}\left( \frac{\kappa_{\rm NN}}{6}+\frac{\kappa_{\rm 3N}}{6}+\kappa_{\rm 4N}+\frac{2\kappa_{\rm 5N}}{3}\right)$} 
    \end{tabular}
    \caption{Free energy changes for FCC lattice up to 4th nearest neighbour bonds (labelled ${\rm 5N}$). Note, the table contains entries for to half of all bonds attached to a single site to avoid double counting. The other half are connected via inversion symmetry and provide identical contributions. We set $\kappa_{\rm 5N}=\kappa_{\rm 4N}$ throughout the paper. Note, we use the shorthand 4N for next-next-nearest neighbours, 3N for next-nearest neighbours and NN for nearest neighbours thoughout.}
    \label{tab:freeenchange}
\end{table*}

\section{Residual Sum-of-Squares}
\label{sec:appendix_rss_contours}

\begin{figure*}[h]
\centering
{
\begin{subfigure}{\textwidth}
\subcaption{}
\includegraphics[width=0.8\textwidth]{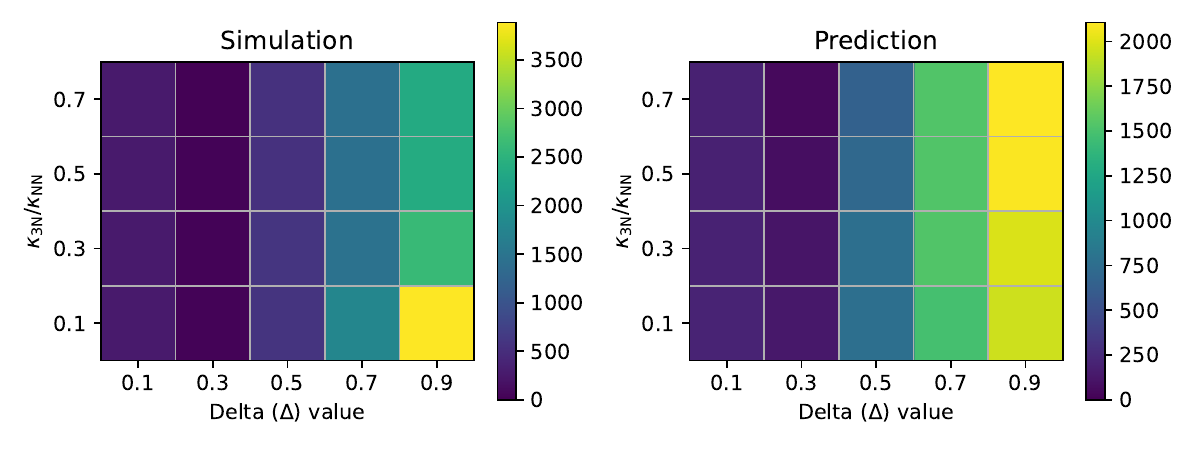}
\end{subfigure}
\begin{subfigure}{\textwidth}
\subcaption{}
\includegraphics[width=0.8\textwidth]{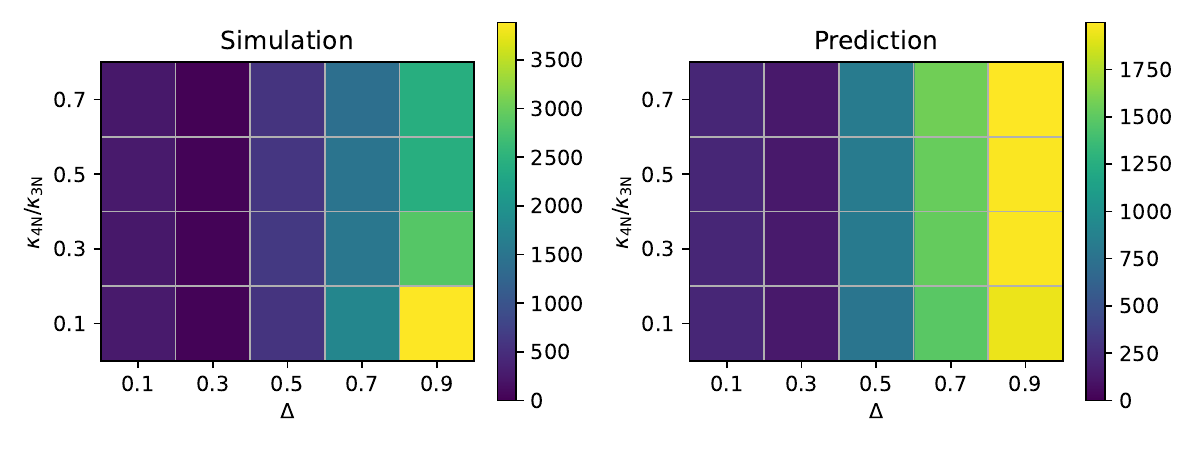}
\end{subfigure}
}
\caption{Variation of the residual sum of squares for glial tissue G1 with (a) $\Delta$ and $\kappa_{\rm 3N}/\kappa_{\rm NN}$, with $\kappa_{\rm 4N}/\kappa_{\rm 3N} = 0.1$; (b) varying parameters $\Delta$ and $\kappa_{\rm 4N}/\kappa_{\rm 3N}$, with $\kappa_{\rm 3N}/\kappa_{\rm NN} = 0.1$. Grids for other tissues are very similar and not shown.}
\label{fig:rss_contour_x1171}
\end{figure*}

\begin{figure*}
{
\begin{subfigure}{\textwidth}
\subcaption{}
\includegraphics[width=0.8\textwidth]{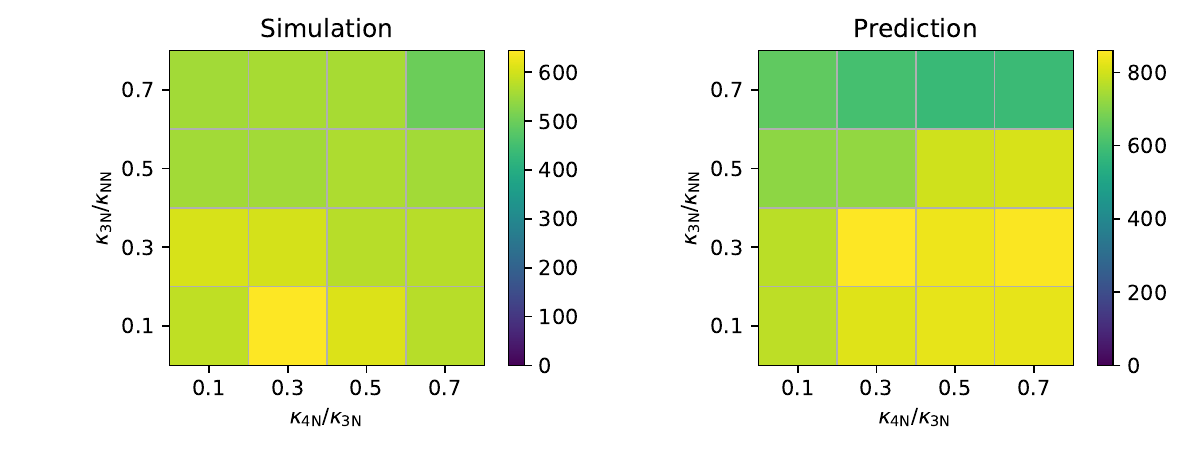}
\end{subfigure}
\begin{subfigure}{\textwidth}
\subcaption{}
\includegraphics[width=0.8\textwidth]{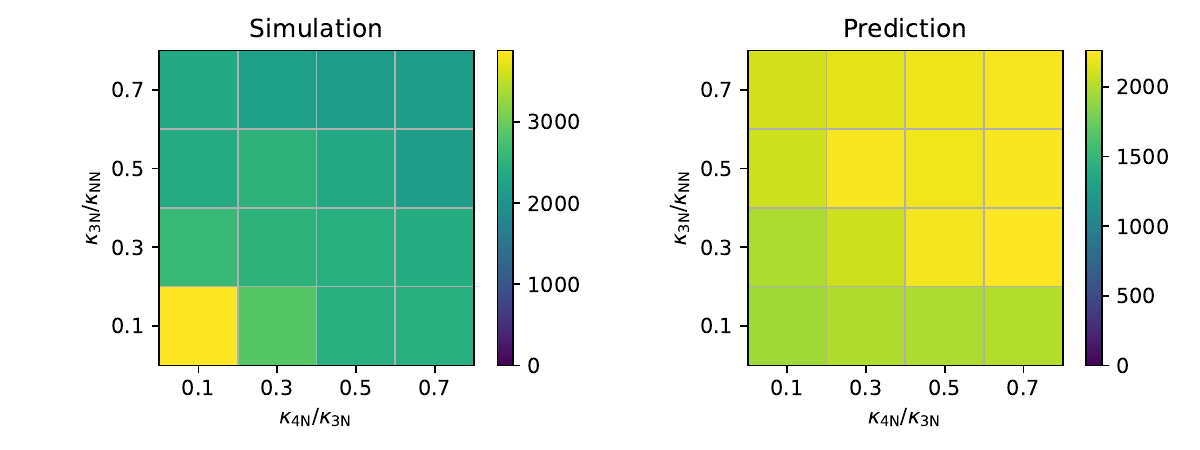}
\end{subfigure}
\caption{Variation of the residual sum of squares for simulations (CONDOR) and predictions (RAPTOR) of glial tissue G1 with (a) $\kappa_{\rm 4N}/\kappa_{\rm 3N}$ and $\kappa_{\rm 3N}/\kappa_{\rm NN}$ with fixed $\Delta = 0.5$; (b) $\kappa_{\rm 4N}/\kappa_{\rm 3N}$ and $\kappa_{\rm 3N}/\kappa_{\rm NN}$ with fixed $\Delta = 0.9$. Grids for other tissues are very similar and not shown.
}
\label{fig:rss_contour_x1171_delfixed}
}
\end{figure*}

As discussed in Section~\ref{sec:validation}, one method used to identify optimal simulation parameters in CONDOR was via a grid search. Corresponding RSS values for the G1 mould are shown in Figures~\ref{fig:rss_contour_x1171} and \ref{fig:rss_contour_x1171_delfixed}. RSS values depend most strongly on $\Delta$.

\section{Pix2pix architecture for RAPTOR}
\label{sec:appendixpix2pix}

Tables \ref{tab:generator_model} and \ref{tab:discriminator_model} summarise the specific architecture of the pix2pix cGAN used in this article, since it differs slightly from Ref. \cite{andrews2023} due to the inclusion of three additional input layers to represent CONDOR model parameters (this leads to a difference in the number of parameters associated with the first 2D convolution layer). All other aspects of the cGAN including the general architecture are the same.

\topcaption{Architecture for the \texttt{pix2pix} discriminator network. Each row in the table corresponds to a single operation in the network. Type describes the names used by TensorFlow to define the network, and output shape the shape of the numerical array that results from operating on the input. For the work in this paper this batch size is always one. Parameter count indicates the numbers of trainable weights corresponding to each layer. The number of parameters in layer 2 differs from that in Ref. \cite{andrews2023} due to the additional 3 input layers that represent the CONDOR model parameters. }\label{tab:generator_model}
{\onecolumn
\begin{mpxtabular*}{\linewidth}{lllr}
\hline
Layer & Type & Output Shape & Parameter Count \\
\hline
1 & InputLayer & (Batch Size, 256, 256, 5) & 0 \\
\hline
2 & Conv2D & (Batch Size, 128, 128, 64) & 5120 \\
3 & LeakyReLU & (Batch Size, 128, 128, 64) & 0 \\
\hline
4 & Conv2D & (Batch Size, 64, 64, 128) & 131072 \\
5 & BatchNormalization & (Batch Size, 64, 64, 128) & 512 \\
6 & LeakyReLU & (Batch Size, 64, 64, 128) & 0 \\
\hline
7 & Conv2D & (Batch Size, 32, 32, 256) & 524288 \\
8 & BatchNormalization & (Batch Size, 32, 32, 256) & 1024 \\
9 & LeakyReLU & (Batch Size, 32, 32, 256) & 0 \\
\hline
10 & Conv2D & (Batch Size, 16, 16, 512) & 2097152 \\
11 & BatchNormalization & (Batch Size, 16, 16, 512) & 2048 \\
12 & LeakyReLU & (Batch Size, 16, 16, 512) & 0 \\
\hline
13 & Conv2D & (Batch Size, 8, 8, 512) & 4194304 \\
14 & BatchNormalization & (Batch Size, 8, 8, 512) & 2048 \\
15 & LeakyReLU & (Batch Size, 8, 8, 512) & 0 \\
\hline
16 & Conv2D & (Batch Size, 4, 4, 512) & 4194304 \\
17 & BatchNormalization & (Batch Size, 4, 4, 512) & 2048 \\
18 & LeakyReLU & (Batch Size, 4, 4, 512) & 0 \\
\hline
19 & Conv2D & (Batch Size, 2, 2, 512) & 4194304 \\
20 & BatchNormalization & (Batch Size, 2, 2, 512) & 2048 \\
21 & LeakyReLU & (Batch Size, 2, 2, 512) & 0 \\
\hline
22 & Conv2D & (Batch Size, 1, 1, 512) & 4194304 \\
23 & BatchNormalization & (Batch Size, 1, 1, 512) & 2048 \\
24 & LeakyReLU & (Batch Size, 1, 1, 512) & 0 \\
\hline
25 & Conv2DTranspose & (Batch Size, 2, 2, 512) & 4194304 \\
26 & BatchNormalization & (Batch Size, 2, 2, 512) & 2048 \\
27 & Dropout & (Batch Size, 2, 2, 512) & 0 \\
28 & ReLU & (Batch Size, 2, 2, 512) & 0 \\
\hline
29 & Concatenate with 21 & (Batch Size, 2, 2, 1024) & 0 \\
\hline
30 & Conv2DTranspose & (Batch Size, 4, 4, 512) & 8388608 \\
31 & BatchNormalization & (Batch Size, 4, 4, 512) & 2048 \\
32 & Dropout & (Batch Size, 4, 4, 512) & 0 \\
33 & ReLU & (Batch Size, 4, 4, 512) & 0 \\
\hline
34 & Concatenate with 18 & (Batch Size, 4, 4, 1024) & 0 \\
\hline
35 & Conv2DTranspose & (Batch Size, 8, 8, 512) & 8388608 \\
36 & BatchNormalization & (Batch Size, 8, 8, 512) & 2048 \\
37 & Dropout & (Batch Size, 8, 8, 512) & 0 \\
38 & ReLU & (Batch Size, 8, 8, 512) & 0 \\
\hline
39 & Concatenate with 15 & (Batch Size, 8, 8, 1024) & 0 \\
\hline
40 & Conv2DTranspose & (Batch Size, 16, 16, 512) & 8388608 \\
41 & BatchNormalization & (Batch Size, 16, 16, 512) & 2048 \\
42 & ReLU & (Batch Size, 16, 16, 512) & 0 \\
\hline
43 & Concatenate with 12 & (Batch Size, 16, 16, 1024) & 0 \\
\hline
44 & Conv2DTranspose & (Batch Size, 32, 32, 256) & 4194304 \\
45 & BatchNormalization & (Batch Size, 32, 32, 256) & 1024 \\
46 & ReLU & (Batch Size, 32, 32, 256) & 0 \\
\hline
47 & Concatenate with 9 & (Batch Size, 32, 32, 512) & 0 \\
\hline
48 & Conv2DTranspose & (Batch Size, 64, 64, 128) & 1048576 \\
49 & BatchNormalization & (Batch Size, 64, 64, 128) & 512 \\
50 & ReLU & (Batch Size, 64, 64, 128) & 0 \\
\hline
51 & Concatenate with 6 & (Batch Size, 64, 64, 256) & 0 \\
\hline
52 & Conv2DTranspose & (Batch Size, 128, 128, 64) & 262144 \\
53 & BatchNormalization & (Batch Size, 128, 128, 64) & 256 \\
54 & ReLU & (Batch Size, 128, 128, 64) & 0 \\
\hline
55 & Concatenate with 3 & (Batch Size, 128, 128, 128) & 0 \\
\hline
56 & Conv2DTranspose & (Batch Size, 256, 256, 8) & 16392 \\
\hline
\hline
Total Parameters & & & 54435080 \\
\hline
\end{mpxtabular*}
\twocolumn}

\begin{table*}
    \centering
    \caption{Architecture for the \texttt{pix2pix} discriminator network. Each row in the table corresponds to a single operation in the network. Type describes the names used by TensorFlow to define the network, and output shape the shape of the numerical array that results from operating on the input. For the work in this paper this batch size is always one. Parameter count indicates the numbers of trainable weights corresponding to each layer. The number of parameters in layer 4 differs from that in Ref. \cite{andrews2023} due to the additional 3 input layers that represent the CONDOR model parameters.}
    \label{tab:discriminator_model}
\begin{tabular}{lllr}
Layer & Type & Output Shape & Parameter Count \\
\hline
1 & InputLayer (Mould Image) & (Batch Size, 256, 256, 5) & 0 \\
2 & InputLayer (Label or Gen. Image) & (Batch Size, 256, 256, 8) & 0 \\
\hline
3 & Concatenate 1 and 2 & (Batch Size, 256, 256, 10) & 0 \\
\hline
4 & Conv2D & (Batch Size, 128, 128, 64) & 13312 \\
5 & LeakyReLU & (Batch Size, 128, 128, 64) & 0 \\
\hline
6 & Conv2D & (Batch Size, 64, 64, 128) & 131072 \\
7 & BatchNormalization & (Batch Size, 64, 64, 128) & 512 \\
8 & LeakyReLU & (Batch Size, 64, 64, 128) & 0 \\
\hline
9 & Conv2D & (Batch Size, 32, 32, 256) & 524288 \\
10 & BatchNormalization & (Batch Size, 32, 32, 256) & 1024 \\
11 & LeakyReLU & (Batch Size, 32, 32, 256) & 0 \\
\hline
12 & ZeroPadding2D & (Batch Size, 34, 34, 256) & 0 \\
13 & Conv2D & (Batch Size, 31, 31, 512) & 2097152 \\
14 & BatchNormalization & (Batch Size, 31, 31, 512) & 2048 \\
15 & LeakyReLU & (Batch Size, 31, 31, 512) & 0 \\
\hline
16 & ZeroPadding2D & (Batch Size, 33, 33, 512) & 0 \\
17 & Conv2D & (Batch Size, 30, 30, 1) & 8193 \\
\hline
\hline
Total Parameters & & & 2774529 \\
\hline
\end{tabular}
\end{table*}

\end{document}